\let\includefigures=\iftrue
%
\let\useblackboard=\iftrue
%
%
\newfam\black
\input harvmac
\noblackbox
\includefigures
\message{If you do not have epsf.tex (to include figures),}
\message{change the option at the top of the tex file.}
\input epsf
\def\figin{\epsfcheck\figin}\def\figins{\epsfcheck\figins}
\def\epsfcheck{\ifx\epsfbox\UnDeFiNeD
\message{(NO epsf.tex, FIGURES WILL BE IGNORED)}
\gdef\figin##1{\vskip2in}\gdef\figins##1{\hskip.5in}
\else\message{(FIGURES WILL BE INCLUDED)}%
\gdef\figin##1{##1}\gdef\figins##1{##1}\fi}
\def\DefWarn#1{}
\def\figinsert{\goodbreak\midinsert}
\def\ifig#1#2#3{\DefWarn#1\xdef#1{fig.~\the\figno}
\writedef{#1\leftbracket fig.\noexpand~\the\figno}%
\figinsert\figin{\centerline{#3}}\medskip\centerline{\vbox{
\baselineskip12pt\advance\hsize by -1truein
\noindent\footnotefont{\bf Fig.~\the\figno:} #2}}
\bigskip\endinsert\global\advance\figno by1}
\else
\def\ifig#1#2#3{\xdef#1{fig.~\the\figno}
\writedef{#1\leftbracket fig.\noexpand~\the\figno}%
\global\advance\figno by1}
\fi
%

\def\smallfig#1#2#3{\DefWarn#1\xdef#1{fig.~\the\figno}
\writedef{#1\leftbracket fig.\noexpand~\the\figno}%
\figinsert\figin{\centerline{#3}}\medskip\centerline{\vbox{
\baselineskip12pt\advance\hsize by -1truein
\noindent\footnotefont{\bf Fig.~\the\figno:} #2}}
\endinsert\global\advance\figno by1}

\useblackboard
\message{If you do not have msbm (blackboard bold) fonts,}
\message{change the option at the top of the tex file.}
\font\blackboard=msbm10 scaled \magstep1
\font\blackboards=msbm7
\font\blackboardss=msbm5
\textfont\black=\blackboard
\scriptfont\black=\blackboards
\scriptscriptfont\black=\blackboardss

\else

\fi
%



\def\boxit#1{\vbox{\hrule\hbox{\vrule\kern8pt
\vbox{\hbox{\kern8pt}\hbox{\vbox{#1}}\hbox{\kern8pt}}
\kern8pt\vrule}\hrule}}
\def\mathboxit#1{\vbox{\hrule\hbox{\vrule\kern8pt\vbox{\kern8pt
\hbox{$\displaystyle #1$}\kern8pt}\kern8pt\vrule}\hrule}}

\def\subsubsection#1{\bigskip\noindent
{\it #1}}

\def\yboxit#1#2{\vbox{\hrule height #1 \hbox{\vrule width #1
\vbox{#2}\vrule width #1 }\hrule height #1 }}
\def\fillbox#1{\hbox to #1{\vbox to #1{\vfil}\hfil}}
\def\ybox{{\lower 1.3pt \yboxit{0.4pt}{\fillbox{8pt}}\hskip-0.2pt}}
%
%


\def\bphi{{\bar \phi}}

\def\bthet{\bar \theta }

\def\bi{{\bar i}}
\def\jb{{\bar j}}
\def\bz{{\bar z}}

\def\tPhi{{\tilde{\Phi}}}
\def\tphi{{\tilde{\phi}}}
\def\hthet{\hat{\theta}}
\def\bhthet{\hat{\bthet}}

\def\l{\left}

\def\comments#1{}

\def\p{\partial}

\def\half{{1\over 2}}
\def\Tr{{{\rm Tr~ }}}
\def\tr{{\rm tr\ }}

\def\bra#1{{\langle}#1|}
\def\ket#1{|#1\rangle}

\def\vev#1{\langle{#1}\rangle}

\def\CA{{\cal A}}
\def\CC{{\cal C}}

\def\CF{{\cal F}}

\def\CN{{\cal N}}
\def\CO{{\cal O}}

\def\CL{{\cal L}}

\def\CS{{\cal S}}
\def\CW{{\cal W}}

\def\II{\relax{I\kern-.10em I}}

\font\cmss=cmss10 \font\cmsss=cmss10 at 7pt
\def\IZ{\relax\ifmmode\mathchoice
{\hbox{\cmss Z\kern-.4em Z}}{\hbox{\cmss Z\kern-.4em Z}}
{\lower.9pt\hbox{\cmsss Z\kern-.4em Z}}
{\lower1.2pt\hbox{\cmsss Z\kern-.4em Z}}
\else{\cmss Z\kern-.4emZ}\fi}
\def\IR{\relax{\rm I\kern-.18em R}}
\def\IZ{\relax\ifmmode\mathchoice
{\hbox{\cmss Z\kern-.4em Z}}{\hbox{\cmss Z\kern-.4em Z}}
{\lower.9pt\hbox{\cmsss Z\kern-.4em Z}} {\lower1.2pt\hbox{\cmsss
Z\kern-.4em Z}}\else{\cmss Z\kern-.4em Z}\fi}
\def\IB{\relax{\rm I\kern-.18em B}}
\def\IC{{\relax\hbox{$\inbar\kern-.3em{\rm C}$}}}
\def\ID{\relax{\rm I\kern-.18em D}}
\def\IE{\relax{\rm I\kern-.18em E}}
\def\IF{\relax{\rm I\kern-.18em F}}
\def\IG{\relax\hbox{$\inbar\kern-.3em{\rm G}$}}
\def\IGa{\relax\hbox{${\rm I}\kern-.18em\Gamma$}}
\def\IH{\relax{\rm I\kern-.18em H}}
\def\II{\relax{\rm I\kern-.18em I}}
\def\IK{\relax{\rm I\kern-.18em K}}
\def\IP{\relax{\rm I\kern-.18em P}}

%

\def\jb{{\bar \jmath}}

\def\inbar{\,\vrule height1.5ex width.4pt depth0pt}

\def\p{\partial}

\font\cmss=cmss10 
\def\IR{\relax{\rm I\kern-.18em R}}

\def\wb{{\bar{w}}}

%


%

\def\lp10{\ell_p^{10}}
\def\lp11{\ell_p^{11}}
\def\R11{R_{11}}

\def\frac#1#2{{#1 \over #2}}

\def\bi{\bar i}
\def\jb{\bar j}

\def\bD{{\bar \Delta}}

\def\l{\left}

\def\comments#1{}

\def\p{\partial}

\def\half{{1\over 2}}
\def\Tr{{{\rm Tr~ }}}
\def\tr{{\rm tr\ }}

\def\bra#1{{\langle}#1|}
\def\ket#1{|#1\rangle}

\def\vev#1{\langle{#1}\rangle}

\def\CA{{\cal A}}
\def\CC{{\cal C}}

\def\CF{{\cal F}}

\def\CH{{\cal H}}

\def\CN{{\cal N}}
\def\CO{{\cal O}}

\def\CL{{\cal L}}

\def\CS{{\cal S}}
\def\CW{{\cal W}}


\def\cf{{\it c.f.}}
\def\M4{M_{Pl,4}}

\def\k11{\kappa_{11}}
\def\l11{\ell_{11}}
\def\tl11{\tilde{\ell}_{11}}

\def\m11{M_{11}}
\def\tm11{\tilde{M}_{11}}

\def\np{{\it Nucl. Phys.}}
\def\prl{{\it Phys. Rev. Lett.}}
\def\pr{{\it Phys. Rev.}}
\def\pl{{\it Phys. Lett.}}

\def\cqg{{\it Class. Quant. Grav.}}

\def\cmp{{\it Comm. Math. Phys.}}

\def\jhep{{\it J. High Energy Phys.}}

\def\etal{{\it et.\ al.}}
\def\eg{{\it e.g.}}

\lref\dv{R. Dijkgraaf and C. Vafa, 
\np\ {\bf B644} (2002) 3 [arXiv:hep-th/0206255];
R, Dijkgraaf and C. Vafa, 
\np\ {\bf B644} (2002) 21
[arXiv:hep-th/0207106];
R. Dijkgraaf and C. Vafa,
[arXiv:hep-th/0208048].}
\lref\OoguriGX{
H.~Ooguri and C.~Vafa,
\np\ {\bf B641} (2002) 3
[arXiv:hep-th/0205297].}

\lref\wb{J. Wess and J. Bagger, {\it Supersymmetry and
Supergravity} (2nd. ed.), Princeton University Press,
Princeton, NJ , 1992.}
\lref\amatietal{D. Amati, K. Konishi, Y. Meurice, G.C. Rossi,
and G. Veneziano, 
{\it Phys. Rep.} {\bf 162} (1988) 169.} 
\lref\nsvza{V.A. Novikov, M.A. Shifman, A.L. Vainshtein,
and V.I. Zakharov, 
\np\ {\bf B229} (1983) 407.}
\lref\rv{G.C. Rossi and G. Veneziano, 
\pl\ {\bf B138} (1983) 195.}
\lref\grza{M.~T.~Grisaru and D.~Zanon,
\np\ {\bf B252} (1985) 578.}

\lref\SeibergRS{
N.~Seiberg and E.~Witten,
\np\ {\bf B426} (1994) 19,
[Erratum-ibid.\ {\bf B430} (1994) 485]
[arXiv:hep-th/9407087].
}

\lref\vy{G. Veneziano and S. Yankielowicz,
\pl\ {\bf B113} (1982) 231.}
\lref\pesk{T. Taylor, G. Veneziano and S.
Yankielowicz, 
\np\ {\bf B218} (1983) 493;
M. Peskin in {\it Problems in 
Unification and Supergravity}, G. Farrar
and F. Heney, eds, AIP, New York, 1984.}

\lref\DijkgraafPP{
R.~Dijkgraaf, S.~Gukov, V.~A.~Kazakov and C.~Vafa,
arXiv:hep-th/0210238.}
\lref\greevy{J. McGreevy, 
[arXiv:hep-th/0211109].}
\lref\dbflat{D. Berenstein, 
\pl\ {\bf B552} (2003) 255
[arXiv:hep-th/0210183].}
\lref\multitrace{V. Balasubramanian, J. de Boer, 
B. Feng, Y.-H. He, M.-x. Huang, V. Jejjala, and
A. Naqvi, 
[arXiv:hep-th/0212082].}
\lref\flavah{R.~Argurio, V.~L.~Campos, G.~Ferretti and R.~Heise,
[arXiv:hep-th/0210291]; H.~Suzuki,
[arXiv:hep-th/0211052]; I.~Bena and R.~Roiban,
\pl\ {\bf B555} (2003) 117
[arXiv:hep-th/0211075];
Y.~Demasure and R.~A.~Janik,
\pl\ {\bf B553} (2003) 105
[arXiv:hep-th/0211082]; 
Y.~Tachikawa,
[arXiv:hep-th/0211189]; 
B.~Feng and Y.~H.~He,
[arXiv:hep-th/0211234];
R.~Argurio, V.~L.~Campos, G.~Ferretti and R.~Heise,
\pl\ {\bf B553} (2003) 332
[arXiv:hep-th/0211249];
S.~G.~Naculich, H.~J.~Schnitzer and N.~Wyllard,
\jhep\ {\bf 0301} (2003) 015
[arXiv:hep-th/0211254];
I.~Bena, R.~Roiban and R.~Tatar,
[arXiv:hep-th/0211271]; B.~Feng,
[arXiv:hep-th/0211202]; K.~Ohta,
\jhep\ {\bf 0302} (2003) 057
[arXiv:hep-th/0212025];
C.~Hofman, 
[arXiv:hep-th/0212095];
N.~Seiberg, ``Adding fundamental matter 
to 'Chiral rings and anomalies in  supersymmetric gauge theory',''
\jhep\ {\bf 0301} (2003) 061
[arXiv:hep-th/0212225];
B.~Feng, 
[arXiv:hep-th/0212274].}
\lref\brandetal{
A.~Brandhuber, H.~Ita, H.~Nieder, Y.~Oz and C.~Romelsberger,
[arXiv:hep-th/0303001].}

\lref\dvpert{R. Dijkgraaf, M.T. Grisaru, C.S. Lam, C. Vafa,
and D. Zanon, 
[arXiv:hep-th/0211017].}
\lref\cdswone{F. Cachazo, M.R. Douglas, N. Seiberg
and E. Witten, 
\jhep\ {\bf 0212} (2002) 071
[arXiv:hep-th/0211170].}

\lref\bipz{E. Brezin, C. Itzykson, G. Parisi, and
J.B. Zuber, 
\cmp\ {\bf 59} (1978) 35.}

\lref\oraf{L. O'Raifeartaigh, 
\np\ {\bf B96} (1975) 331.}
\lref\constraints{E. Witten, 
\np\ {\bf B202} (1982) 253.}
\lref\adsa{I. Affleck, M. Dine, and N. Seiberg, 
\pl\ {\bf B137}
(1984) 187.}
\lref\adsb{I. Affleck, M. Dine, and N. Seiberg, 
\prl\ {\bf 52} (1984) 1677.}
\lref\adsc{I. Affleck, M. Dine, and N. Seiberg, 
\np\ {\bf B241} (1984) 493.}
\lref\adsd{I. Affleck, M. Dine, and N. Seiberg,
\np\ {\bf B256} (1985) 557.}
\lref\iss{K. Intriligator, N. Seiberg, and S. Shenker,
\pl\ {\bf B342} (1995) [arXiv:hep-th/9410203].}
\lref\izya{K. Izawa and T. Yanagida, 
{\it Prog.\ Theor.\ Phys.}  {\bf 95} (1996) 829
[arXiv:hep-th/9602180];
K. Intriligator and S. Thomas, 
\np\ {\bf B473} (1996) 121
[arXiv:hep-th/9603158];
K. Intriligator and S. Thomas,
[arXiv:hep-th/9608046].}
\lref\intha{
K. Intriligator and S. Thomas, 
\np\ {\bf B473} (1996) 121
[arXiv:hep-th/9603158]
.}
\lref\inthb{
K. Intriligator and S. Thomas,
[arXiv:hep-th/9608046].}
\lref\konvez{K. Konishi and G. Veneziano, 
\pl\  {\bf B187} (1987) 106.}
\lref\ggsoft{L. Girardello and M.T. Grisaru,
\np\ {\bf B194} (1982) 65.}
\lref\EvansIA{
N.~Evans, S.~D.~Hsu and M.~Schwetz,
\pl\ {\bf B355} (1995) 475
[arXiv:hep-th/9503186];
N.~Evans, S.~D.~H.~Hsu, M.~Schwetz and S.~B.~Selipsky,
Nucl.\ Phys.\ B {\bf 456}, 205 (1995)
[arXiv:hep-th/9508002];
N.~Evans, S.~D.~H.~Hsu and M.~Schwetz,
Phys.\ Lett.\ B {\bf 404}, 77 (1997)
[arXiv:hep-th/9703197].
}

\lref\marcos{
L.~Alvarez-Gaume and M.~Marino,
arXiv:hep-th/9606168;
L.~Alvarez-Gaume and M.~Marino,
Int.\ J.\ Mod.\ Phys.\ A {\bf 12}, 975 (1997)
[arXiv:hep-th/9606191].
}

\lref\dn{M. Dine and A. Nelson, 
\pr\ {\bf D48} (1993) 1277,
[arXiv:hep-ph/9303230].}
\lref\dns{M. Dine, A. Nelson, and Y. Shirman,
\pr\ {\bf D51} (1995) 1362 [arXiv:hep-ph/9408384].}
\lref\dnns{M. Dine, A. Nelson, Y. Nir, and Y. Shirman,
\pr\ {\bf D53} (1996) 2658
[arXiv:hep-ph/9507378].}

\lref\dysonPT{F. Dyson, 
\pr\ {\bf 85} (12962) 31.}

\lref\WittenEP{
E.~Witten,
Nucl.\ Phys.\ B {\bf 507}, 658 (1997)
[arXiv:hep-th/9706109].
}
\lref\bdlr{
I.~Brunner, M.~R.~Douglas, A.~Lawrence and C.~Romelsberger,
\jhep\ {\bf 0008} (2000) 015
[arXiv:hep-th/9906200].}
\lref\dougstrings{M.R. Douglas, ``Mastering
$N=1$,'' talk given at Strings 2002 in Cambridge,
UK.  See the transparencies at
{\tt http://www.damtp.cam.ac.uk/strings02/avt/douglas/index.html}.}
\lref\DouglasVM{
M.~R.~Douglas,
\cqg\  {\bf 17} (2000) 1057,
[arXiv:hep-th/9910170];
M.~R.~Douglas, B.~Fiol and C.~Romelsberger,
[arXiv:hep-th/0003263];
D.~E.~Diaconescu and M.~R.~Douglas,
[arXiv:hep-th/0006224];
M.~R.~Douglas, 
[arXiv:math.ag/0009209];
M.~R.~Douglas,
[arXiv:hep-th/0105014].}
\lref\DouglasFR{
M.~R.~Douglas, S.~Govindarajan, T.~Jayaraman and A.~Tomasiello,
[arXiv:hep-th/0203173].}
\lref\DouglasGI{
M.~R.~Douglas,
{\it J.\ Math.\ Phys.}  {\bf 42} (2001) 2818
[arXiv:hep-th/0011017].}

\lref\DouglasFJ{
M.~R.~Douglas,
[arXiv:math.ag/0207021].}
\lref\BerensteinFI{
D.~Berenstein and M.~R.~Douglas,
[arXiv:hep-th/0207027].}

\lref\kklma{S.~Kachru, S.~Katz, A.~Lawrence and J.~McGreevy,
\pr\ {\bf D62} (2000) 026001
[arXiv:hep-th/9912151].}
\lref\reidca{M. Reid, ``Minimal models of canonical 3-folds,''
pp. 131-180, Advanced Studies in Pure Mathematics 1, ed. S. Iitaka,
Kinokuniya (1983).}

\lref\KachruAN{
S.~Kachru, S.~Katz, A.~E.~Lawrence and J.~McGreevy,
\pr\ {\bf D62} (2000) 126005;
[arXiv:hep-th/0006047].
}
\lref\CandelasDM{
P.~Candelas, X.~De La Ossa, A.~Font, S.~Katz and D.~R.~Morrison,
\np\ {\bf B416} (1994) 481,
[arXiv:hep-th/9308083].
}
\lref\KatzHT{
S.~Katz, D.~R.~Morrison and M.~Ronen Plesser,
\np\ {\bf 477} (1996) 105,
[arXiv:hep-th/9601108].
}
\lref\CachazoPR{
F.~Cachazo and C.~Vafa,
arXiv:hep-th/0206017.
}

\lref\GVW{
S.~Gukov, C.~Vafa and E.~Witten,
\np\ {\bf B584} (2000) 69,
[Erratum-{\it ibid.} {\bf B608} (2001) 477],
[arXiv:hep-th/9906070].}
\lref\TaylorII{
T.~R.~Taylor and C.~Vafa,
\pl\ {\bf B474} (2000) 130
[arXiv:hep-th/9912152].}
\lref\CurioSC{
G.~Curio, A.~Klemm, D.~Lust and S.~Theisen,
\np\ {\bf B609} (2001) 3
[arXiv:hep-th/0012213].}
\lref\GranaJJ{
M.~Grana and J.~Polchinski,
\pr\ {\bf D63} (2001) 026001,
[arXiv:hep-th/0009211].
}
\lref\GranaXN{
M.~Grana and J.~Polchinski,
\pr\ {\bf D65} (2002) 126005,
[arXiv:hep-th/0106014].
}
\lref\GubserVG{
S.~S.~Gubser,
arXiv:hep-th/0010010.
}
\lref\GiddingsYU{
S.~B.~Giddings, S.~Kachru and J.~Polchinski,
Phys.\ Rev.\ D {\bf 66}, 106006 (2002)
[arXiv:hep-th/0105097].
}
\lref\vafabraneflux{Vafa's brane-flux duality papers}
\lref\CamaraKU{
P.~G.~Camara, L.~E.~Ibanez and A.~M.~Uranga,
arXiv:hep-th/0311241.
}
\lref\Grananew{
M.~Grana, T.~W.~Grimm, H.~Jockers and J.~Louis,
arXiv:hep-th/0312232.}
\lref\CachazoJY{
F.~Cachazo, K.~A.~Intriligator and C.~Vafa,
Nucl.\ Phys.\ B {\bf 603} (2001) 3,
[arXiv:hep-th/0103067].
}
\lref\PolchinskiUF{
J.~Polchinski and M.~J.~Strassler,
arXiv:hep-th/0003136.
}
\lref\KlebanovNC{
I.~R.~Klebanov and A.~A.~Tseytlin,
Nucl.\ Phys.\ B {\bf 578} (2000) 123, [arXiv:hep-th/0002159].
}
\lref\MaldacenaYY{
J.~M.~Maldacena and C.~Nunez,
Phys.\ Rev.\ Lett.\  {\bf 86}, 588 (2001)
[arXiv:hep-th/0008001].
}
\lref\geometrictransitions{
R.~Gopakumar and C.~Vafa,
{\it Adv.\ Theor.\ Math.\ Phys.}  {\bf 2} (1998) 413,
[arXiv:hep-th/9802016];
R.~Gopakumar and C.~Vafa,
{\it Adv.\ Theor.\ Math.\ Phys.}  {\bf 3} (1999) 1415,
[arXiv:hep-th/9811131];
J.~M.~Maldacena and C.~Nunez,
Phys.\ Rev.\ Lett.\  {\bf 86}, 588 (2001)
[arXiv:hep-th/0008001];
I.~R.~Klebanov and A.~A.~Tseytlin,
Nucl.\ Phys.\ B {\bf 578} (2000) 123, [arXiv:hep-th/0002159];
J.~Polchinski and M.~J.~Strassler,
arXiv:hep-th/0003136.
}
\lref\GopakumarKI{
R.~Gopakumar and C.~Vafa,
{\it Adv.\ Theor.\ Math.\ Phys.}  {\bf 3} (1999) 1415,
[arXiv:hep-th/9811131].
}

\lref\PolchinskiSM{
J.~Polchinski and A.~Strominger,
\pl\ {\bf B388} (1996) 736,
[arXiv:hep-th/9510227].
}
\lref\kklt{
S.~Kachru, R.~Kallosh, A.~Linde and S.~P.~Trivedi,
{\it Phys.\ Rev.} {\bf D68} (2003) 046005,
[arXiv:hep-th/0301240];
S.~Kachru, R.~Kallosh, A.~Linde, J.~Maldacena, 
L.~McAllister and S.~P.~Trivedi,
JCAP {\bf 0310} (2003) 013,
[arXiv:hep-th/0308055].
}
\lref\KachruSX{
S.~Kachru, R.~Kallosh, A.~Linde, J.~Maldacena, 
L.~McAllister and S.~P.~Trivedi,
JCAP {\bf 0310} (2003) 013,
[arXiv:hep-th/0308055].
}
\lref\CopelandBJ{
E.~J.~Copeland, R.~C.~Myers and J.~Polchinski,
arXiv:hep-th/0312067.
}

\lref\KachruNS{
S.~Kachru, X.~Liu, M.~B.~Schulz and S.~P.~Trivedi,
JHEP {\bf 0305} (2003) 014, 
[arXiv:hep-th/0205108].
}
\lref\KachruSK{
S.~Kachru, M.~B.~Schulz, P.~K.~Tripathy and S.~P.~Trivedi,
JHEP {\bf 0303}, 061 (2003)
[arXiv:hep-th/0211182].
}

\lref\nutcharge{
C.~M.~Hull,
Nucl.\ Phys.\ B {\bf 509}, 216 (1998)
[arXiv:hep-th/9705162].
}

\lref\GiryavetsVD{
A.~Giryavets, S.~Kachru, P.~K.~Tripathy and S.~P.~Trivedi,
arXiv:hep-th/0312104.
}

\lref\GreeneDH{
B.~R.~Greene, D.~R.~Morrison and C.~Vafa,
\np\ {\bf B481} (1996) 513,
[arXiv:hep-th/9608039].
}

\lref\DineKV{
M.~Dine and N.~Seiberg,
\prl\ {\bf 55} (1985) 366.
}
\lref\KaplunovskyYY{
V.~S.~Kaplunovsky,
Phys.\ Rev.\ Lett.\  {\bf 55} (1985) 1036.
}
\lref\WittenMZ{
E.~Witten,
\np\ {\bf B471} (1996) 135,
[arXiv:hep-th/9602070].
}
\lref\BanksSS{
T.~Banks and M.~Dine,
\np\ {\bf B479} (1996) 173,
[arXiv:hep-th/9605136].
}
\lref\KaloperUJ{
N.~Kaloper, M.~Kleban, A.~E.~Lawrence and S.~Shenker,
\pr\ {\bf D66} (2002) 123510,
[arXiv:hep-th/0201158].
}

\lref\DimopoulosZB{
S.~Dimopoulos and H.~Georgi,
\np\ {\bf B193} (1981) 150.
}

\lref\mayr{
P.~Mayr,
Nucl.\ Phys.\ B {\bf 593}, 99 (2001)
[arXiv:hep-th/0003198].
}
\lref\bwsb{A. Anisimov, M.~Dine, M.~Graesser, and
S. Thomas, 
[arXiv:hep-th/0111235];
[arXiv:hep-th/hep-th/0201256].}
\lref\granaferm{M.~Gra\~na, 
[arXiv:hep-th/0202118].}
\lref\dewolfescales{O.~DeWolfe and S.~B.~Giddings,
[arXiv:hep-th/0208123].}
\lref\granamssm{M.~Gra\~na, 
[arXiv:hep-th/0209200].}

\lref\anommed{M.~Dine and D.~MacIntire,
\pr\ {\bf D46} (1992) 2594, [arXiv:hep-ph/9205227];
L.~Randall and R.~Sundrum,
\np\ {\bf B557} (1999) 79, [arXiv:hep-th/9810155];
G.~F.~Giudice, M.~A.~Luty, H.~Murayama and R.~Rattazzi,
{\it J. High Energy Phys.} {\bf 9812} (1998) 027
[arXiv:hep-ph/9810442].
}
\lref\ginomed{Z.~Chacko, M.~A.~Luty and E.~Ponton,
JHEP {\bf 0007} (2000) 036,
[arXiv:hep-ph/9909248];
D.~E.~Kaplan, G.~D.~Kribs and M.~Schmaltz,
\pr\ {\bf D62} (2000) 035010,
[arXiv:hep-ph/9911293];
Z.~Chacko, M.~A.~Luty, A.~E.~Nelson and E.~Ponton,
{\it J. High Energy Phys.} {\bf 0001} (2000) 003,
[arXiv:hep-ph/9911323];
M.~Schmaltz and W.~Skiba,
\pr\ {\bf D62} (2000) 095005
[arXiv:hep-ph/0001172];
D.~E.~Kaplan and G.~D.~Kribs,
{\it J. Hight Energy Phys.} {\bf 0009} (2000) 048,
[arXiv:hep-ph/0009195].
}
\lref\dkkls{S.~Dimopoulos, S.~Kachru, N.~Kaloper, 
A.~E.~Lawrence and E.~Silverstein,
\pr\ {\bf D64} (2001) 121702,
[arXiv:hep-th/0104239];
S.~Dimopoulos, S.~Kachru, N.~Kaloper, A.~E.~Lawrence and E.~Silverstein,
arXiv:hep-th/0106128.}

\lref\berksieg{N.~Berkovits and W.~Siegel, 
\np\ {\bf B462} (1996) 213, 
[arXiv:hep-th/9510106].}
\lref\VafaWI{
C.~Vafa,
{\it J. Math. Phys.}  {\bf 42} (2001) 2798,
[arXiv:hep-th/0008142].
}
\lref\BerkovitsPQ{
N.~Berkovits, H.~Ooguri and C.~Vafa,
arXiv:hep-th/0310118.
}
\lref\GrimmExtend{R. Grimm, M. Sohnius, and J. Wess,
{\it Nucl. Phys.} {\bf B133} (1978) 275.}
\lref\DeWitvanH{B. de Wit and J.W. van Holten,
\np\ {\bf B155} (1979) 530.}
\lref\DeRoochiral{M. de Roo, J.W. van Holten, B. de Wit,
and A. van Proeyen, 
\np\ {\bf B173} (1980) 175.}
\lref\AntoniadisVB{
I.~Antoniadis, H.~Partouche and T.~R.~Taylor,
Phys.\ Lett.\ B {\bf 372}, 83 (1996)
[arXiv:hep-th/9512006].
}
\lref\AntoniadisKI{
I.~Antoniadis and T.~R.~Taylor,
{\it Fortsch.\ Phys.}  {\bf 44} (1996),
[arXiv:hep-th/9604062].
}
\lref\PartoucheYP{
H.~Partouche and B.~Pioline,
{\it Nucl.\ Phys.\ Proc.\ Suppl.}  {\bf 56B} (1997) 322
[arXiv:hep-th/9702115].
}

\lref\BerkovitsWR{
N.~Berkovits,
\np\ {\bf B431} (1994) 258,
[arXiv:hep-th/9404162].
}
\lref\BerkovitsTG{
N.~Berkovits, S.~Gukov and B.~C.~Vallilo,
Nucl.\ Phys.\ B {\bf 614}, 195 (2001)
[arXiv:hep-th/0107140].
}

\lref\BurgessHX{
C.~P.~Burgess, E.~Filotas, M.~Klein and F.~Quevedo,
[arXiv:hep-th/0209190].
}
\lref\KleinVU{
M.~Klein,
\pr\ {\bf D66} (2002) 055009,
[arXiv:hep-th/0205300];
C.~P.~Burgess, E.~Filotas, M.~Klein and F.~Quevedo,
[arXiv:hep-th/0209190].
}

\lref\AtickGY{
J.~J.~Atick, L.~J.~Dixon and A.~Sen,
\np\ {\bf B292} (1987) 109.
}
\lref\DineGJ{
M.~Dine, I.~Ichinose and N.~Seiberg,
\np\ {\bf B293} (1987) 253.
}
\lref\CecottiUV{
S.~Cecotti, S.~Ferrara and L.~Girardello,
\pl\ {\bf B206} (1988) 451.
}
\lref\OvrutGK{
B.~A.~Ovrut,
\pl\ {\bf B205} (1988) 455.
}
\lref\BanksCY{
T.~Banks, L.~J.~Dixon, D.~Friedan and E.~J.~Martinec,
\np\ {\bf B299}, 613 (1988).
}
\lref\BanksYZ{
T.~Banks and L.~J.~Dixon,
\np\ {\bf B307} (1998) 93.
}
\lref\BeckerKB{
K.~Becker, M.~Becker and A.~Strominger,
\np\ {\bf B456} (1995) 130,
[arXiv:hep-th/9507158].
}
\lref\OoguriCK{
H.~Ooguri, Y.~Oz and Z.~Yin,
\np\ {\bf B477} (1996) 407,
[arXiv:hep-th/9606112].
}

\lref\HullZY{
C.~M.~Hull,
\np\ {\bf B267} (1986) 266.
}

\lref\BershadskyCX{
M.~Bershadsky, S.~Cecotti, H.~Ooguri and C.~Vafa,
{\it Commun.\ Math.\ Phys.} {\bf 165} (1994) 311,
[arXiv:hep-th/9309140].
}
\lref\BershadskyTA{
M.~Bershadsky, S.~Cecotti, H.~Ooguri and C.~Vafa,
\np\ {\bf B405} (1993) 279,
[arXiv:hep-th/9302103].
}

\lref\CandelasUG{
P.~Candelas, P.~S.~Green and T.~Hubsch,
\np\ {\bf B330} (1990) 49.
}
\lref\CandelasPI{
P.~Candelas and X.~de la Ossa,
\np\ {\bf B355} (1991) 455.
}

\lref\StromingerPD{
A.~Strominger,
\cmp\ {\bf 133} (1990) 163.
}

\lref\joebook{J. Polchinski, {\it String Theory} vol. 1-2, 
Cambridge University Press (1998).}
\lref\gukov{
S.~Gukov,
Nucl.\ Phys.\ B {\bf 574}, 169 (2000)
[arXiv:hep-th/9911011].
}

\lref\softads{
N.~Evans, M.~Petrini and A.~Zaffaroni,
JHEP {\bf 0206}, 004 (2002)
[arXiv:hep-th/0203203];
O.~Aharony, E.~Schreiber and J.~Sonnenschein,
the
JHEP {\bf 0204}, 011 (2002)
[arXiv:hep-th/0201224];
V.~Borokhov and S.~S.~Gubser,
JHEP {\bf 0305}, 034 (2003)
[arXiv:hep-th/0206098];
S.~Kuperstein and J.~Sonnenschein,
arXiv:hep-th/0309011.
}

\lref\SchwarzQR{
J.~H.~Schwarz,
\np\ {\bf B226} (1983) 269.
}
\lref\HassanBV{
S.~F.~Hassan,
\np\ {\bf B568} (2000) 145,
[arXiv:hep-th/9907152].
}

\lref\othersusyb{O.~Aharony, 
J.~Sonnenschein, M.~E.~Peskin and S.~Yankielowicz,
\pr\ {\bf D52} (1995) 6157,
[arXiv:hep-th/9507013];
E.~D'Hoker, Y.~Mimura and N.~Sakai,
\pr\ {\bf D54} (1996) 7724,
[arXiv:hep-th/9603206];
H.~C.~Cheng and Y.~Shadmi,
Nucl.\ Phys.\ B {\bf 531}, 125 (1998)
[arXiv:hep-th/9801146];
N.~Arkani-Hamed and R.~Rattazzi,
Phys.\ Lett.\ B {\bf 454}, 290 (1999)
[arXiv:hep-th/9804068];
M.~A.~Luty and R.~Rattazzi,
theories
JHEP {\bf 9911}, 001 (1999)
[arXiv:hep-th/9908085];
G.~R.~Farrar, G.~Gabadadze and M.~Schwetz,
Phys.\ Rev.\ D {\bf 60}, 035002 (1999)
[arXiv:hep-th/9806204].
}

\lref\FarrarRM{
G.~R.~Farrar, G.~Gabadadze and M.~Schwetz,
Phys.\ Rev.\ D {\bf 60}, 035002 (1999)
[arXiv:hep-th/9806204].
}

\lref\baryonsandbranesinads{
E.~Witten,
JHEP {\bf 9807}, 006 (1998)
[arXiv:hep-th/9805112];
D.~J.~Gross and H.~Ooguri,
Phys.\ Rev.\ D {\bf 58}, 106002 (1998)
[arXiv:hep-th/9805129].
}


\lref\nongeometric{
S.~Hellerman, J.~McGreevy and B.~Williams,
arXiv:hep-th/0208174;
A.~Dabholkar and C.~Hull,
JHEP {\bf 0309}, 054 (2003)
[arXiv:hep-th/0210209];
A. Flournoy, B. Wecht, B. Williams, to appear.
}

\lref\FidanzaZI{
S.~Fidanza, R.~Minasian and A.~Tomasiello,
arXiv:hep-th/0311122.
}

\lref\syz{
A.~Strominger, S.~T.~Yau and E.~Zaslow,
Nucl.\ Phys.\ B {\bf 479}, 243 (1996)
[arXiv:hep-th/9606040].
}

\lref\othernongeometric{other nongeometric refs?}
\lref\simeon{S. Hellerman, Princeton journal club.}

\lref\NSmirror{Discussions
of mirror symmetry in the presence
of NS flux  can be found for example
in: S.~Gurrieri, J.~Louis, A.~Micu and D.~Waldram,
\np\ {\bf B654} (2003) 61 [arXiv:hep-th/0211102]; 
S.~Gurrieri and A.~Micu,
\cqg\ {\bf 20} (2003) 2181 [arXiv:hep-th/0212278];
S.~Fidanza, R.~Minasian and A.~Tomasiello,
arXiv:hep-th/0311122.}

\lref\shaljm{S. Hellerman, A. Lawrence, and J. McGreevy, work in progress.}

\Title{\vbox{\baselineskip12pt\hbox{hep-th/0401034}
\hbox{BRX TH-513} \hbox{PUTP-2104}}}
{\vbox{
\centerline{Local string models of soft supersymmetry breaking}}}
\smallskip
\centerline{Albion Lawrence$^1$ and John McGreevy$^2$}
\bigskip
\centerline{$^{1}${Martin Fisher School of Physics, Brandeis 
University,}}
\centerline{{MS 057, PO Box 549110, Waltham, MA 02454-9110}}
\medskip
\centerline{$^{2}${Dept. of Physics, Princeton University,
Princeton, NJ 08544}}
\bigskip
\bigskip
\noindent

We study soft 
supersymmetry breaking 
in local models of type II string theory compactifications
with branes and fluxes. In such models, magnetic fluxes
can be treated as auxiliary fields in $\CN=2$ SUSY multiplets.
These multiplets appear as
``spurion superfields'' in the low-energy effective action
for the local model. We discuss the 
pattern of SUSY breaking from $\CN=2$ to $\CN=1$ to $\CN=0$ in these
models, and then identify the fields leading to soft
SUSY breaking terms in various examples.  
In the final
section, we reconsider arguments for the Dijkgraaf-Vafa
conjecture in gauge theories with softly broken supersymmetry.

\Date{January 2004}

\newsec{Introduction}

A wide class of string theory backgrounds 
with low-energy $\CN=1$ supersymmetry
in four dimensions is described by
combinations of D-branes, orientifold planes, and magnetic fluxes
in curved compact manifolds.
Gauge dynamics and charged matter will arise from open strings
when D-branes wrap cycles 
and fill the four-dimensional spacetime.  
Interesting physics 
may also arise via D-branes which
wrap small cycles of the manifold and so give rise to
light nonabelian gauge fields and charged matter.

In models consistent with the unification of standard model
couplings at a high scale, various directions of the
internal manifold 
are somewhat
large compared to the 10-dimensional Planck scale
\refs{\DineKV,\KaplunovskyYY,\WittenMZ,\BanksSS,\KaloperUJ}.
The gauge degrees of freedom and the chiral matter will
typically be localized.  Supersymmetry breaking may occur
in some region of the compactification manifold,
distinct from the visible sector, perhaps via strong gauge dynamics. 

If we wish low-energy SUSY in the visible sector
to subdue the hierarchy problem, supersymmetry in the
visible sector must be broken by explicit 
soft terms \refs{\DimopoulosZB}.  In this paper we will focus on the
description of tree-level soft SUSY-breaking parameters
in local models of 
D-branes near singularities.\foot{``D-branes
near singularities'' is meant to be vague: it can refer to
space-filling D-branes placed near singularities,
or light particle states in four dimensions that arise from
Dp-branes wrapping vanishing p-cycles.} 
In such models, there is a useful softly-broken $\CN=2$ structure
arising from the underlying 
$\CN=2$ supersymmetric closed-string theory 
without branes.  From this point of view 
the soft-breaking parameters 
will be described by auxiliary components of
closed string fields, if those fields couple to relevant
operators in the brane Lagrangian.
The auxiliary fields are typically 
magnetic fluxes 
with indices along the Calabi-Yau directions.
This is the string theory realization of the ``spurion'' method
for describing soft SUSY-violating terms \refs{\ggsoft}.

Such a description of the SUSY-breaking vevs 
is of interest for a number of reasons.
\item{1.} 
These models describe a local
piece of a compactification 
of some cosmologically interesting 
compactifications of string theory \refs{\kklt}.
The SUSY-breaking
can happen elsewhere in the compactification manifold,
perhaps via strong gauge dynamics.
Such models 
have led to interesting scenarios such as anomaly mediation
\refs{\anommed}, gaugino mediation \refs{\ginomed}, 
and ``tunneling'' mediation \refs{\dkkls} which use the
physical separation in extra dimensions
of low-energy degrees of freedom in an essential way.  
More generally, these constructions allow 
one to describe physics in a modular fashion --
in various throats of the geometry 
one has a Standard Model module, 
a supersymmetry breaking module, and
possibly an inflation module.
These machines communicate only via closed string modes 
on the CY, which are constrained
at string tree level (at least) by the $\CN=2$ supersymmetry 
discussed here.

\item{2.} Such a description allows one to plug into the Berkovits 
formalism \refs{\berksieg, \BerkovitsWR} for describing the 
string worldsheet physics for Calabi-Yau compactifications.  
In this description, the vertex operators come 
in spacetime supermultiplets, 
and the expectation values for RR fields do not 
present an obstruction.  This formalism was
partial inspiration for this paper, but we will leave
an explicit discussion of it for future work.

\item{3.}
With such a description of SUSY breaking, the
underlying supersymmetry is still apparent and
one may attempt to make recourse to $\CN=1$, $d=4$
SUSY nonrenormalization theorems to compute
terms in the effective lagrangian.  This has been done
in the past in the field theory context, as in
\refs{\EvansIA, \othersusyb}.  This approach is known
to have limitations, particularly if the mass scale set
by the soft terms is on the order of or larger than
other dynamical scales of the theory.  At the end of this
paper, we will reconsider the Dijkgraaf-Vafa proposal \dv\
in the presence of explicit soft SUSY-breaking terms.

\vskip .2cm

Previous work on describing soft SUSY-breaking terms in
4d string models has been done
in \refs{\mayr,\bwsb,\granaferm,\dewolfescales,\granamssm}\foot{While 
this work was nearing 
completion, \refs{\CamaraKU, \Grananew} appeared.
These papers approach the question of 
soft supersymmetry breaking 
on the worldvolumes of D3-branes 
from a complementary supergravity perspective.}
for a subset of SUSY-breaking terms or for fairly 
specific models.  
Soft SUSY-breaking terms in the near-horizon limit 
are studied in \softads.
Soft SUSY-breaking in $\CN=2$ gauge 
theory was studied using the Seiberg-Witten solution
in \marcos.
In this paper we will discuss 
these terms more generally within the context of type IIB
compactifications, keeping track of
the pattern of SUSY breaking via the
``spurion'' approach, and 
making contact with the string worldsheet.

The outline of this paper is as follows.  In \S2 we will
discuss the superspace description of
$\CN=2$ multiplets for Calabi-Yau moduli
in type II string compactifications, 
incorporating 
the discussion of \refs{\berksieg},
and identifying explicitly the auxiliary fields
in terms of fields in $10d$ supergravity.
This fleshes out (and modifies some details of) the
discussion in \refs{\VafaWI}.
In \S3 we will discuss the explicit breaking
of supersymmetry, for gauge dynamics
realized either by open strings or by wrapped
D-branes. This will include a review
of Vafa's derivation \refs{\VafaWI} of the superpotential
for complex stucture moduli \refs{\GVW,\TaylorII}. 
In \S4 we will reconsider the
Dijkgraaf-Vafa proposal in the light of explicit
soft SUSY-breaking terms.  In \S5 we 
confront our limitations and present
conclusions and possibilities for future work.

\newsec{Auxiliary fields for closed string modes}

In local type II models describing D-branes in a noncompact 
Calabi-Yau background, the closed string 
fields live in $\CN=2$, $d=4$ supermultiplets.
Their vevs determine the coupling constants 
on the worldvolumes of the branes.  
The underlying $\CN=2$ supersymmetry is nonlinearly realized
on the D-brane worldvolumes.
$\CN=2$ supersymmetry is also nonlinearly realized
in local models with
magnetic fluxes, as Vafa and collaborators
have explained \refs{\VafaWI-\BerkovitsPQ}.  
These two statements are related, as branes can 
be transmogrified into fluxes 
by variations of closed string moduli.  
A particularly striking example of this 
is the near-horizon limit
\refs{\eg\ \geometrictransitions, \CachazoJY},
as the closed string moduli are driven through 
topology-changing geometric
transitions.

The $\CN=2$, $d=4$ 
transformations of the closed string multiplets
constrain the manner in which they 
appear in the D-brane effective action.  For example, 
``decoupling theorems'' in 
\refs{\bdlr,\DouglasVM}
state that to all orders in perturbation theory,
closed string hypermultiplets do not couple to the superpotential
for open string chiral scalar multiplets.\foot{The 
claim of these references
is not that all F-terms for open string degrees
of freedom are independent of
closed string hypermultiplets.  In particular, the
gauge coupling for open string vector multiplets
is also an F-term.  At tree level, the string theory computation
is not topological, and for wrapped B(A)-type D-branes
in IIB(A), the gauge coupling clearly depends on the volume of
the cycle the brane is wrapping, which lies in a hypermultiplet.
Note also that at one loop order, the 
open string gauge coupling is topological
\refs{\BershadskyCX}, 
but suffers from a 
holomorphic anomaly \refs{\BershadskyTA}.}
With this in mind, it seems important to
understand soft supersymmetry breaking starting with
the underlying $\CN=2$ structure of closed string
degrees of freedom.\foot{Recent work has
been done on theories with nonlinearly realized
$\CN=2$ supersymmetry in \refs{\KleinVU};
it would be useful to fit the general 
discussion in \refs{\KleinVU}\
into our framework, but we leave this for future work.}


In this section we will provide a complete
identification of $\CN=2$ auxiliary fields and 
closed-string fluxes, using the
$\CN=2$ superspace description of the massless closed string
multiplets provided in \refs{\GrimmExtend,\DeWitvanH,\DeRoochiral,
\berksieg}, by deriving the map between
the auxiliary components of the supermultiplets and magnetic fluxes.
This superspace formalism is natural
from the worldsheet point of view.  In either
the RNS or the Berkovits-Siegel formalisms,
the $\CN=2$ supersymmetry algebra is built from
a $\CN=1$ subalgebra that can be constructed
from a left-moving worldsheet current, and a
$\CN=1$ subalgebra that can be constructed from
a right-moving worldsheet current.
From the spacetime point of view, one has two copies of
$\CN=1$ superspace variables, $(\theta^\alpha, \bthet^{\dot{\alpha}})$ 
and $(\hthet^\alpha, \bhthet^{\dot{\alpha}})$.
In the Berkovits-Siegel formalism, these
appear explicitly as anticommuting fields on the worldsheet
paired by spacetime supersymmetry to the
$4d$ uncompactified target space coordinates.
The left-moving supercharges are constructed from the
unhatted superspace variables, and the right-moving
supercharges are constructed from the superspace
variables with hats.  The
superspace variables form a doublet 
$(\theta^{\alpha},\hthet^{\alpha})$ under the
$SU(2)$ R-symmetry.

Although the discussion in \refs{\berksieg}
takes place within the ``hybrid formalism'', which is related
to the RNS description by a complicated field redefinition
\refs{\BerkovitsWR}, the field redefinition is chiral, so that
the identification of the superspace coordinates with left- and
right-moving supersymmetry currents allows us to understand
vertex operators for the auxiliary fields in the RNS formalism,
using the techniques of \refs{\AtickGY,\DineGJ}.
The worldsheet currents for spacetime supersymmetry
contain the spectral flow operators for the internal
$c=9$ $\CN=(2,2)$ SCFT, together with spin fields
for the $4d$ spacetime coordinates.
Therefore, if the bottom component
of the superfield is an NS-NS field, then
one may identify the coefficients of
$\theta^2$ or $\hat{\theta}^2$ in the superspace
expansion with NS-NS states, and
the coefficients of $\theta \hat{\theta}$ with RR states.

We will discuss in turn the superfield 
description of the vector- and hypermultiplets for type II
compactifications on Calabi-Yau backgrounds.  More precisely,
we will discuss vector multiplets in type IIB string theory
and hypermultiplets in type IIA string theory.  For vectors
in IIA and hypers in IIB, some of the auxiliary fields will be
what are loosely called
``mirrors of NS flux'' \refs{\VafaWI,\KachruSK,\NSmirror}.  
These latter cases will be explored in future work \refs{\shaljm}.
We will close the section with a brief discussion of the
relationship to the hybrid formalism of \refs{\BerkovitsWR,\berksieg}
to our discussion, to address points where our results appear
to disagree.

\subsec{Massless vector multiplets}

In type IIB string theory compactified
on a Calabi-Yau threefold $X$, the 
scalar component of the massless vector multiplets
are complex structure deformations of the Calabi-Yau 
background.  One may write the full vector multiplet 
in the $\CN=2$ superspace language of 
\refs{\GrimmExtend,\DeWitvanH,\DeRoochiral,\berksieg}.
We will discuss here the chiral superfield,
which satisfies the constraint \refs{\GrimmExtend,
\DeRoochiral}:~\foot{Here $\sigma^{\mu} = (1,\vec{\sigma})$ is a four-vector
of $2\times 2$ matrices, where $\vec{\sigma}$ are the
Pauli matrices and $1$ denotes the identity.  
Furthermore, $\bar{\sigma}^{\mu} = (1,-\vec{\sigma})$,
and 
$\sigma^{\mu\nu} = \frac{1}{4}\left[\sigma^{\mu}\bar{\sigma}^{\nu} -
\sigma^{\nu}\bar{\sigma}^{\mu}\right]$.}
\eqn\chiralconstraint{
\eqalign{
   \bar{\nabla}_{\dot{\alpha}} V & \equiv 
   \left(- \frac{\p}{\p\bar{\theta}^{\dot{\alpha}}} 
      - i \sigma^\mu_{\beta \dot{\alpha}}
      \theta^{\beta} \p_{\mu}\right) V = 0 \cr
   \hat{\bar{\nabla}}_{\dot{\alpha}} V & \equiv \left(- 
      \frac{\p}{\p\hat{\bar{\theta}}^{\dot{\alpha}}} 
      - i \sigma^\mu_{\beta \dot{\alpha}}
      \hat{\theta}^{\beta} \p_{\mu}\right) V = 0\ .
}}
The superspace expansion for $V$ is:
\eqn\twovector{
\eqalign{
   V & ~=~ w^a + \theta^\alpha \zeta^a_\alpha+\hthet^\alpha\hat{\zeta}^a_\alpha
      + \theta^2 D_{++}^a + 
      \theta^\alpha\hthet^\beta
      \left(\epsilon_{\alpha\beta}D_{+-}^a + F^a_{\alpha\beta}\right)
      \cr
       & \ \ + \hthet^2 D^a_{--}
      + \theta^\alpha\hthet^2 \chi^a_{\alpha} +
      \hthet^\beta\theta^2
      \hat{\chi}^a_{\beta}\cr
      &\ \ \ + \theta^2 \hthet^2 C^a\ .
}}
Here $w^a$ denotes the complex structure deformation.
$D_{ij}$ is a symmetric tensor in the $SU(2)$ indices
$i,j$, with complex entries, and is an auxiliary field.  
$C$ is also a complex auxiliary field. 
Finally,
$$ F_{\mu\nu} = \sigma^{\alpha\beta}_{\mu\nu} F_{\alpha\beta} $$ 
is an anti-self-dual antisymmetric tensor. 
This has $16+16$ bosonic plus fermionic coordinates.  

One may apply further superspace constraints to cut the number
of off-shell degrees of freedom in half 
\refs{\GrimmExtend,\DeWitvanH,\DeRoochiral,\berksieg},
using the superspace constraint
\eqn\supercon{
   (\epsilon_{ij}\nabla^i \sigma_{\mu\nu} \nabla^j)
   (\epsilon_{kl}\nabla^k \sigma^{\mu\nu} \nabla^l) V
   = - 96 \p^2 \bar{V}
}
Here $i$ is an $SU(2)$ doublet index and $\nabla^i = 
(\nabla,\hat{\nabla})$. In component form the contraints imply:

\item{1.} $\p^2 D_{++} = \p^2 D^*_{--}$, and
$\p^2 D_{+-}$ is real.  Note that this
places no constraints on constant modes of $D_{ij}$ \refs{\PartoucheYP}.

\item{2.}  $F^a_{\alpha\beta}$ satisfies the identities
\eqn\fsident{
   \sigma_\mu^{\alpha\dot{\alpha}} \p_\mu F^a_{\alpha\beta}
   = \sigma^\mu_{\beta\dot{\beta}} \p_\mu F^{a,\dot{\alpha}\dot{\beta}}
}
and so is an anti-self-dual abelian vector field strength.

\item{3.} $\chi_{\alpha} = \sigma^{\mu}_{\alpha\dot{\alpha}}
\p_{\mu}\bar{\zeta}^{\dot{\alpha}}$, and
$\hat{\chi}_{\alpha} = \sigma^{\mu}_{\alpha\dot{\alpha}}
\p_{\mu}\hat{\bar{\zeta}}^{\dot{\alpha}}$.

\item{4.} $C^a = \p^2 \bar{w}^a$

These components comprise one complex scalar, a
vector field, an $SU(2)$ triplet of auxiliary fields,
and an $SU(2)$ doublet of Weyl fermions.  Note that since
$\theta,\hthet$ correspond respectively to the left- and right-moving
supercharges on the worldsheet, $D_{+-}$ is a Ramond-Ramond scalar
and $F_{\alpha\beta}$ is a Ramond-Ramond vector, while
$w$ and $D_{\pm\pm}$ are NS-NS scalars.

Although this multiplet is clearly simpler, we will
work with the less constrained chiral multiplet,
letting BRST invariance take care of the
reduction to on-shell degrees of freedom.  There
are at least two reasons that this seems advantageous.  First,
we will find that the auxiliary fields 
of the unconstrained multiplet are naturally
identified with elements of Dolbeault cohomology
in the Calabi-Yau.  Secondly, while the chiral constraints
\chiralconstraint\ are linear in derivatives,
\supercon\ is nonlinear in derivatives.
Thus, while the product of two
chiral fields is a chiral field, the product of two
fields satisfying \supercon\ no longer satisfies
\supercon.

\bigskip
\noindent{\it Auxiliary fields in type IIB on a CY}
\medskip

We wish to begin by stating our results for the identification
of auxiliary fields in $\CN=2$ vector supermultiplets.
In order to state them, it is helpful to review
the various descriptions of the moduli space of complex structures,
which is the vector multiplet moduli space in type IIB string theory
(and is part of the hypermultiplet moduli space in type IIA
string theory). We follow the discussion in
\refs{\CandelasUG,\CandelasPI,\StromingerPD,\BershadskyCX}.
In the large-volume conformal field theory, the natural
description of small deformations of the complex structure is
in terms of metric perturbations of the form
\eqn\csdefmet{
   \delta ds^2 = \delta g_{\bi \jb} d\bz^{\bi} d\bz^{\jb}\ .
}
These small deformations modulo reparametrizations 
correspond to 
vector-valued holomorphic one-forms
\eqn\vvoneform{
   \delta g_{\bi\jb}g^{i\bi}d\bar{z}^{\jb} =
   v^i_{\bar{j}}d\bar{z}^{\bar{j}} \ \in \ H_{\bar \del}^{(0,1)}(X,TX)\ .
}
This group is isomorphic to the Dolbeault cohomology
group $H^{(2,1)}(X)$ by the formula
\eqn\dolbeault{
   \delta g_{\bi\jb}g^{i\jb}\Omega_{ijk} d\bar{z}^{\bi}dz^j dz^k
   = \omega_{\bi j k} d\bar{z}^{\bi}dz^j dz^k\ .
}
The complex dimension of the moduli space is the
complex dimension $b^{21}$ of $H^{(2,1)}(X)$.
We will refer to these coordinates as ``CFT coordinates''.

Another natural set of coordinates on moduli space is
the periods of the holomorphic $(3,0)$ form
$\Omega$.  Let $A^a, B_b, a = 1 \dots b^{21}+1$ 
be a symplectic basis of $H_3(X)$, where
\eqn\intersect{
   A^a \cap B_b = \delta^a_b\ ;\ \ \ \ \
   B_a \cap A^b = - \delta^b_a\ .
}
One may define a dual integral basis $\alpha^a, \beta_b$
for $H^3(X)$, such that
\eqn\dualrelations{
   \int_{A^a} \alpha_b = \int_X \alpha_b \wedge \beta^a = \delta_b^a
   \ ;\ \ \ \ \
   \int_{B_a} \beta^b = \int_X \alpha_a \wedge \beta^b = - \delta_a^b
   \ ;\ \ \ \ \ 
   \int_{A^a} \beta^b = \int_{B_a} \alpha_b = 0\ .
}
In this basis, the $b^{21} + 1$ complex periods
\eqn\periodvars{
   t^a = \int_{A^a} \Omega 
}
form a set of projective coordinates on moduli space. These
are projective because the theory is invariant under
rescalings 
\eqn\recalings{
   \Omega(t) \longrightarrow e^{f(t)} \Omega(t)
}
where $t$ are any set of holomorphic coordinates on
the moduli space of complex structures.  The dual periods
\eqn\dualperiodvars{
   F_a(t) = \int_{B_a} \Omega
}
are determined by $t$.  Alternatively, we could have picked $F$
as the projective coordinates on moduli space and $t$ as the 
dual variables. Locally in moduli space, we can write
\eqn\prepot{
   F = \half \sum_a t^a F_a(t)\ ;\ \ \ \ F_a = \frac{\p}{\p t^a} F
}
$F$ is the prepotential for the vector multiplets, and
has projective weight two.

If we specify an element of $H^{(2,1)}$ by
\eqn\csdolb{
   \omega = \sum_{m=1}^{b_{21}} t^m \omega_m
}
where $\omega_m$ are basis forms for $H^{(2,1)}(X)$,
$t^m$, $m=(1,\ldots,b_{21})$ 
are the coordinates in the coordinate system
specifying an element of $H^{(2,1)}$, we can write
$\Omega = \Omega(t^m)$, and
\eqn\perderiv{
   \frac{\p}{\p t^m} \Omega = k_m \Omega + \omega_m
}
$k_m$ can be shifted by a projective transformation.
We can thereby 
choose a gauge where it vanishes \refs{\StromingerPD}.

Using these coordinates we can now describe the auxiliary
fields.  Two ingredients are the NS-NS three-form $H_{ij\bar{k}}$,
and the RR three-form $F_{ij\bar{k}}$.  The third ingredient
is built from the almost complex structure.  For $\CN=2$
Calabi-Yau vacua of type II string theory, the
complex structure $J_{\mu}^{\nu}$ can be written
as a two-form by lowering the vector index with the metric:
\eqn\csform{
   J = J_{\mu\nu}dx^{\mu} dx^{\nu} = 
   J_{\mu}^{\lambda}g_{\lambda\nu} dx^{\mu}dx^{\nu}\ ,
}
which can be rewritten in complex coordinates:
\eqn\csformcplx{
   J = J_{i\jb} dz^i d\bar{z}^{\jb} \equiv i g_{i\jb}
   dz^i d\bar{z}^{\jb}\ .
}
For $\CN=2$ vacua, $dJ = \p J = \bar{\p} J = 0$.
We will find that NS auxiliary fields are
related to nonvanishing
\eqn\metricaux{
   T = i ( \p -\bar \p ) J .
}
%

In the ``CFT coordinates'', we can expand
\eqn\fluxescft{
\eqalign{
   H & = \sum_m^{b^{21}} h^m \omega_m + H^{(3,0)}
+ {\rm h.c.}\equiv\tilde{H} + H^{(3,0)} +{\rm h.c.}\cr
   F & = \sum_m f^m \omega_m + F^{(3,0)}
+ {\rm h.c.}\equiv\tilde{F}+F^{(3,0)}+{\rm h.c.}\cr
   T & = \sum_m \tau^m \omega_m +T^{(3,0)}
+ {\rm h.c.}\equiv\tilde{T}+T^{(3,0)}+{\rm h.c.}
}}
Here we have defined $\tilde G$ to indicate the 
$(2,1)$ part of the three-form $G$.
We will identify
\eqn\cftauxiliary{
\eqalign{
   D_{++}^m & = \left(\tau^m + h^m\right) \cr
   D_{+-}^m & = g_s (f^m - C^{(0)} h^m)\cr
   D_{--}^m & = \left(\tau^m - h^m\right)\ .
}}
where $C^{(0)}$ is the IIB RR axion.
In particular, the auxiliary fields in
the chiral vector multiplets are $(2,1)$ forms 
made from the fluxes and torsion form.
In the remainder of this section 
we will use worldsheet techniques to justify this claim.

Note that this identification does not 
include the $(3,0)$ piece of $H,F,T$.
In a compact IIB model, 
the relation 
\eqn\momentmap{
  i\int_X \Omega\wedge \bar \Omega = { 4\over 3} \int 
J \wedge J \wedge J }
implies that the deformation of 
$\Omega$ proportional to itself
changes the volume of the CY, 
and hence lies in a hypermultiplet.
In such a compact model, 
this deformation is related by 
spacetime supersymmetry to 
the flux along this direction, 
and hence such flux is the 
auxiliary field in a hypermultiplet.
In a noncompact model, the both hand side 
of \momentmap\ is infinite, 
and the volume deformation does not exist.

\bigskip
\noindent{\it Vertex operators for auxiliary fields}
\medskip

The vertex operators for the auxiliary fields can be derived
using the techniques in \refs{\AtickGY,\DineGJ}, as we shall do here.
Since we do not know how to include a nonzero vev for 
the $RR$ scalar $C^{(0)}$ in the RNS formalism, we will set
$C^{(0)} = 0$ in these discussions.  We will eventually 
find the correct dependence of the auxiliary fields on $C^{(0)}$ via
spacetime arguments.
To begin, let us consider the fields $w^a, \zeta^a, D_{++}^a$, which
form a chiral multiplet under the left-moving supersymmetry.
In spacetime, the supersymmetry transformations of this 
chiral multiplet are:
\eqn\spacetimetrans{
\eqalign{
    \left[ Q_\alpha, w^m\right] & = \zeta^m_\alpha \cr
    \left[ Q_{\dot{\alpha}}, w^m\right] &= 0 \cr
    \left\{ Q_{\alpha}, \zeta^m_{\beta} \right\} &= 
       \epsilon_{\alpha\beta} D^m_{++} \cr
    \left\{ Q_{\dot{\alpha}}, \zeta^m_{\beta} \right\}
       & = i\dsl_{\dot{\alpha}\beta} w^m \cr
    \left[ Q_\alpha, D^m_{++} \right] &= 0\cr
    \left[ Q_{\dot{\alpha}}, D^m_{++} \right] 
       &= i\dsl_{\dot{\alpha}\beta} \zeta^{\beta,m}\ .
}}
Since worldsheet correlators lead to spacetime S-matrix elements, 
the worldsheet vertex operators are
the spacetime fields times the inverse propagator.
The result is that acting on the vertex operator for
$D_{++}$ twice with the left-moving
supersymmetry charge, one gets the vertex operator for
the scalar $w$ without any momentum factors.
One may use this to guess the vertex operator for $D_{++}$.
Following \refs{\AtickGY,\DineGJ}, the
vertex operator for $w$ in the $(-1,0)$ picture
is:~\foot{\cf\ ref. \refs{\joebook} for discussions
of pictures and of the superconformal ghosts.}
\eqn\scalarrnsvo{
   V_w^{(-1,0)} = e^{-\phi} V_{w, CY} \sim e^{-\phi}\delta g_{\bi\jb}
      \bar{\psi}^{\bi}\hat{\p}\bar{X}^{\jb}\ ,
}
where the final expression is the approximate expression at large
radius and complex structure.  Here left-moving fermions and
derivatives are denoted by unhatted variables, and
right-moving fermions and derivatives are denoted by hatted variables.
A physical metric variation $\delta g$ may be expressed in terms 
of a {\it harmonic} $(2,1)$ form 
$ \omega_{\bi jk} = 
\delta g_{\bi\jb} g^{l \jb} \Omega_{ljk}$.

The vertex operator for $D_{++}$ can be
written in the $(0,0)$ picture as:
%
\eqn\vertexaux{
    V_{D_{++}}(z') = \lim_{z\to z'} \left\{
    (z-w) \epsilon^+(z) V_{w,CY}(z') \right\} + \ldots\ .
}
The additional pieces are terms in the vertex operator
which have nonsingular OPEs with the supercurrents.  These
will be necessary for $V_D$ to be physical.
$\epsilon^+$ is the operator generating one unit of spectral
flow from the NS sector back to the NS sector, and at large
volume and complex structure it can be written \AtickGY:
\eqn\specflow{
   \epsilon^+(z) \sim \Omega_{ijk}\psi^i \psi^j \psi^k
}
where $\Omega$ is the holomorphic $(3,0)$ form defining the
complex structure of $X$.

Thus, in the $(0,0)$ picture, at large volume and
complex structure, we can write:
\eqn\auxscvertex{
   V_{D_{++}} = \delta g_{\bi\jb} g^{i\jb} \Omega_{ijk} 
   \psi^j \psi^k \hat{\p}\bar{X}^{\bi} =
   \omega_{\bi jk}\psi^j\psi^k\hat{\p}\bar{X}^{\bi} + \ldots\ ;
}
a label $m = 1 \dots b^{21}$ on $V_D$, $\delta g$ and $\omega$ 
is implied.
This implies \cftauxiliary\ when $C^{(0)} = 0$.\foot{Similar results
for the auxiliary fields in the gravity multiplet
of $\CN=1$ heterotic compactifications can be found
in \refs{\OvrutGK,\CecottiUV}.} In general, terms of the
form $\psi^2 \hat{\p} X$ arise from NS-NS flux and
from the couplings of the worldsheet fermions
to the spacetime affine connection.  The linear combination
in \cftauxiliary\ is precisely that needed to reproduce
the particular holomorphic index structure shown in
\auxscvertex.  
We will describe this relation in more detail 
below when we discuss $D_{\pm\pm}$ in terms of $\sigma$-model couplings.
The arguments are essentially identical for $D_{--}$.

Next we consider the RR auxiliary field $D_{+-}$.  
The action of the spacetime supersymmetry charges 
on the vertex operator for $D_{+-}$ is:
\eqn\susycommrr{
  \left\{ \hat{Q}^{(-1/2)}_{\dot{\alpha}}, 
  \left[Q^{(-1/2)}_{\dot{\beta}}, 
     V^{(-1/2,-1/2)}_{D_{+-}} \right]
  \right\} = \epsilon_{\dot{\alpha}\dot{\beta}} V_w^{(-1,-1)}\ ,
}
where the superscripts denote the picture with respect to
the gauged $\CN=1$ worldsheet superconformal algebra.
One may check that, at large volume and complex structure, 
the following operator has the correct transformation
properties:
\eqn\rrauxvotwo{
   V_{D_{+-}}^{(-\half, -\half)} = 
   g_s e^{-\frac{\phi}{2}-\frac{\hat{\phi}}{2}}
   \epsilon^{\alpha\beta} 
   S_{\alpha}\hat{S}_{\beta} S^{-,\dot{a}} 
   (C\Gamma)^{\bi j k}_{\dot{a}\dot{b}}\hat{S}^{-,\dot{b}}
   \omega_{\bi j k}+\ldots\ .
}
Here $S_\alpha$ is the positive chirality spin field
for the $4d$ spacetime CFT; $S^{-,\dot{a}}$ is a negative chirality
$6d$ spinor transforming as a $\bar{{\bf 4}}$ 
under the $SO(6)$ acting locally
on the tangent space to $X$; $\Gamma^{ABC\ldots}$ are 
antisymmetrized products of the
6d $\Gamma$-matrices with indices in ${\bf 4} \oplus \bar{{\bf 4}}$;
and $C$ is a charge conjugation matrix intertwining the
${\bf 4}$ and $\bar{\bf 4}$ representations.

The vertex operator in \rrauxvotwo\ is clearly $g_s$ times 
the vertex operator for a harmonic RR 3-form field
strength $F \propto \omega$; this verifies
\cftauxiliary\ when $C^{(0)} = 0$.\foot{Recall
that in string frame, the spacetime action for the RR fields
is independent of $g_s$, so the corresponding vertex
operator should also be independent of $g_s$.}   
The additional factor of $g_s$ is needed for $V_w$ to have the 
right normalization.

We should make some cautionary remarks at this point.  
One may only make small deformations of magnetic fields if there are
noncompact three-cycles on which these fields have support.
Otherwise, the fluxes satisfy quantization conditions, and
small deformations that are constant in spacetime 
are not on-shell modes.  (Although, as in \refs{\GVW}, one
may have solutions which interpolate in four dimensions
between different values of magnetic 3-form flux.)
Furthermore, in the case where the flux threads noncompact cycles, one
must still take care with the vertex operators.  In general
they may have logarithmic OPEs with themselves, due to their
behavior at infinity in field space.  
A cautionary example in this regard is the open string 
vertex operator for a constant magnetic field strength 
on a bosonic D-brane in
flat space:~\foot{We thank M. Berkooz for
pointing this out to us.}
\eqn\constantmagfield{
   V = B_{\mu\nu}X^{\mu}\p X^{\nu} + \ldots\ .
}
The term shown is not quite a scaling operator.
Note that such pieces are missing from the analysis above,
as they have nonsingular OPEs with the spacetime supercharges.

\bigskip
\noindent{\it 
Auxiliary fields as sigma model couplings}

Another way to understand the physics of the auxiliary fields
is to ask, in worldsheet language, what sigma model couplings
will break $\CN=2$ spacetime SUSY to $\CN=1$ spacetime SUSY,
and which $\CN=1$ subgroup will be preserved.  Of course,
in the RNS formalism, we will be restricted to considering
$D_{++}$ and $D_{--}$ as we do not know how to treat
nontrivial Ramond-Ramond backgrounds in the RNS formalism.

For example, if the auxiliary field $D_{--}$ has an expectation value,
the supersymmetry corresponding to $\hthet$ is broken;
in particular, $\hat{\zeta}$ becomes a goldstino
for this supersymmetry, 
\eqn\goldsttrans{
   \hat{\delta} \hat{\zeta}_\alpha = \hat{\epsilon}_\alpha D_{--}
}
Therefore, a vev for $D_{--}$ will break the $\CN=1$ supersymmetries
arising from the right-moving sector of the worldsheet,
but preserve the left-moving sector.  The worldsheet theory
must have $\CN=(2,1)$ supersymmetry; the $\CN=2$ supersymmetry
for the left-movers leads to an $\CN=1$ spacetime supersymmetry
\refs{\BanksCY,\BanksYZ}, and the $(0,1)$ part of the worldsheet 
supersymmetry is 
gauged.  Similarly, if we break the
supersymmetries arising from the left-moving supercharges,
the worldsheet theory should have $\CN=(1,2)$ supersymmetry.

General sigma models with $\CN=(2,1)$ supersymmetry 
were described by Hull \refs{\HullZY}.  These models
contain non-vanishing 3-form field strengths which couple
to the worldsheet fermions.  Such field strengths couple to the
left- and right-moving fermions with opposite sign: the
quadratic terms in the Lagrangian are:
\eqn\quadferm{
   L = - i \psi_{\pm}^{\mu} \left[
      g_{\mu\nu} \p_\mp \psi^{\nu}_{\pm} \pm 
      \left( g_{\mu\nu} \Gamma^{\nu}_{\lambda\rho}
      \pm H_{\mu\nu\rho}\right)
      \p_\mp \phi^{\lambda} \psi_{\pm}^{\rho}\right]
}
$\CN=2$ worldsheet supersymmetry in either the left- or right-moving
sector requires a complex structure covariant with respect to the
connection 
\eqn\connection{
   \gamma^{\mu}_{\nu\rho,\pm} = \Gamma^{\mu}_{\nu\rho}
      \pm H^{\mu}_{\nu\rho}
}
The result is that the metric $g$ should 
be Hermitian, 
the three-form field strength should be a $(2,1)$-form, and
that the metric and NS-NS field strength should satisfy the equations:
\eqn\partialsusy{
   H_{ij\bar{k}} = \pm i \left(\p_j g_{i\bar{k}} - 
   \p_i g_{j\bar{k}}\right) = \pm T_{\bi jk}\ .
}
This is consistent  with \cftauxiliary.  If $H = T$, then
the left-moving supersymmetry is broken and the
right-moving supersymmetry is intact, since 
$D_{++} \neq 0$, $D_{--} = 0$.  If we choose the
opposite sign in \partialsusy, the roles of 
$D_{++}$, $D_{--}$ are reversed.

\bigskip
\noindent{\it Auxiliary fields for periods of fluxes}
\medskip


If one chooses the moduli space coordinates to
be the periods $t^a$, then the auxiliary 
components of the resulting supermultiplets are:
\eqn\periodauxiliary{
\eqalign{
   D_{++}^a & = \int_{A^a} \left(\tilde{T} + \tilde{H}\right) \cr
   D_{+-}^a & = g_s \int_{A^a} \left(\tilde{F}-C^{(0)}\tilde{H}\right)\cr
   D_{--}^a & = \int_{A^a} \left(\tilde{T} - \tilde{H}\right)\ .
}}
Eq. \periodauxiliary\ can be demonstrated using \cftauxiliary\
as follows.  (Here we have restored the dependence on
$C^{(0)}$, as we will justify below.)
One can write the periods in terms of CFT coordinates,
as implied by eq. \perderiv.
%
Now if $V$ is an $\CN=2$ chiral superfield of the form
\twovector, $t^a(V)$ is also a chiral superfield.
Its lowest component is just the period $t^a(w)$.
To find the auxiliary field, we need merely
expand $t^a(V)$ to second order in $\theta, \hat{\theta}$.
Let us compute the coefficient $D_{++}^a$ of $\theta^2$.
We get this by expanding $t$ to first order in
the auxiliary fields $D_{++}^m$ for $w^m$:
\eqn\newauxfromold{
   D_{++}^a = D_{++}^m \frac{\p t^a}{\p w^m}
}
Now, using \perderiv, and choosing a gauge such that
$k_m = 0$ at the point in moduli space of interest, we find:
\eqn\newauxfinally{
   D_{++}^a = \sum_m \int_{A^a} (\tau^m + h^m) \omega_m
   = \int_{A^a} (\tilde{T} + \tilde{H})\ .
}
The computation for the other auxiliary fields
is nearly identical, so \periodauxiliary\ 
indeed follows from \cftauxiliary.
If one chooses instead the dual coordinates
$F_a$, then an analogous computation shows that
\eqn\dualperiodauxiliary{
\eqalign{
   D_{++, a} & = \int_{B_a} \left(\tilde{T}+\tilde{H}\right) \cr
   D_{+-, a} & = g_s \int_{B_a} \left(\tilde{F}-C^{(0)}\tilde{H}\right) \cr
   D_{--, a} & = \int_{B_a} \left(\tilde{T} - \tilde{H}\right)
}}
It will be useful in \S3 to observe that 
these facts \periodauxiliary\dualperiodauxiliary\
can be summarized by a 'three-form superfield' 
of the form
\eqn\threeformsuperfield{\Omega(\theta,\hat \theta) 
= \Omega + \theta^2 \left (\tilde T + \tilde H \right)
+ \hat \theta^2 \left (\tilde T - \tilde H \right)
+ \theta \hat \theta \left (\tilde F - C^{(0)} \tilde H \right)
+ \dots.}

\bigskip
\noindent{\it D-branes and auxiliary fields}

While it is difficult to discuss vevs for RR
fields from the worldsheet in RNS language, we
can examine spacetime solutions which have RR flux
and $\CN=1$ supersymmetry, in order to 
confirm our identifications
\cftauxiliary,\periodauxiliary,\dualperiodauxiliary.

Let us start with a large number of 
space-filling D-branes in type IIB
string theory, wrapped on a holomorphic cycle
so as to preserve $\CN=1$ supersymmetry in four dimensions.
Such D-branes preserve 
the supersymmetry generated by 
$Q_\alpha + e^{ i \gamma}\hat{Q}_\alpha$
for some phase $e^{i \gamma}$.
For a single brane $\gamma$ may be set to unity by 
redefining the phase of $\hat Q$.
But if other branes and fluxes are present which
also by themselves break $\CN=2$ SUSY to an 
$\CN=1$ subalgebra
defined by a different phase,
the relative phases will matter.

Brane-flux duality 
\refs{\geometrictransitions-\CachazoJY}
states that the string background with
D-branes is dual to a geometric background with nontrivial 
Ramond-Ramond flux.  
A particular solution describing
D5-branes wrapped around a holomorphic 2-cycle is
given in \refs{\MaldacenaYY}
when $C^{(0)} = 0$.  
Using this
fact, the authors of \refs{\VafaWI,\BerkovitsPQ}\ have argued that
one should set $D_{++}$, $D_{+-}$, and $D_{--}$ all equal
to $F$ in order to preserve the same spacetime supersymmetry
as the D-brane.  This entails preserving the supersymmetry
$Q + \hat{Q}$, which is precisely the supersymmetry
preserved by a D5-brane wrapping a holomorphic cycle
at large volume.\foot{The boundary conditions on the fermions are
$\psi = - \hat{\psi}$ for worldsheet superpartners
to Dirichlet directions and
$\psi = \hat{\psi}$ for superpartners to Neumann directions.
One may use this to deduce the action of
the boundary conditions on the spectral flow operator,
which at large volume can be constructed from the
RNS fermions by bosonization.}
These authors identify $D_{\pm,\pm}$ with NS-NS flux.
Our identification of auxiliary fields
implies that $H=0$, $T = g_s F$.\foot{This value of $T$ can
be inferred from S-duality: the S-dual NS-NS solution on
\refs{\MaldacenaYY} must satisfy $H = \pm T$
in order to preserve $\CN=1$ SUSY, as we have discussed earlier.}
This identification of NS-NS auxiliary fields
solves a slight puzzle in
comparing the discussion of \refs{\VafaWI,\BerkovitsPQ}\
to the spacetime solution in \refs{\MaldacenaYY}, as
the solution in \refs{\MaldacenaYY}\ contains no
NS-NS magnetic flux: the relevant NS-NS field is the
three-form $T$ built from the complex structure.

The S-dual solution  in \MaldacenaYY\ describes an
$NS5$-brane wrapping the same cycle.  In this case,
we can still preserve four supercharges, but there is
no source for an RR field.  Such solutions should
therefore correspond to either $D_{++}$ or $D_{--}$ being nonzero,
while the other vanishes.  Which vanishes depends on whether
one wraps a fivebrane or an anti-fivebrane around the cycle.
The anti-fivebrane is a source for 3-form flux with sign opposite
to the flux generated by a fivebrane.  Therefore, if
the Maldacena-Nu\~nez solution for a fivebrane is such that
$H = T$, leading to $D_{--} = 0$, the solution for
an anti-fivebrane has $-H = T$, leading to $D_{++} = 0$.

We can appeal to spacetime arguments to ask
about the auxiliary fields when $C^{(0)}$ is nonvanishing.
The simplest argument for the appearance of the $C^{(0)}$-dependent
terms in \cftauxiliary,\periodauxiliary, and
\dualperiodauxiliary\ is to notice how the spacetime
spinors transform with $F$ in type IIB \refs{\SchwarzQR,\HassanBV}.
One can see directly from the string-frame presentation in
Appendix B of \refs{\HassanBV}\ that $F$ always appears in the
combination $F - C^{(0)} H$.

Finally, in compact models with branes and fluxes, one can turn
on a combination of NS-NS and RR fluxes to maintain supersymmetry
in the presence of nonvanishing D3-branes.  The supersymmetry
preserved by D3-branes is $Q + i \hat{Q}$.  If we set
$T = 0$, we can expand the vector multiplets out in 
powers of $\theta \pm i \hat{\theta}$.  Choosing coordinates
equal to periods $t$ of the CY, we find:
\eqn\newexp{
   V = t + ig_s (\theta + i\hat{\theta})^2 \tilde{G} -
   i g_s(\theta - i \hat{\theta})^2 \tilde{G}^* + \ldots
}
where $\tilde{G} = \tilde{F} - \tau \tilde{H}$, 
and $\tau = C^{(0)} + \frac{i}{g_s}$, and the
tildes denote the components of the forms lying in
$H^{(2,1)}(X)$.  Note that
if $G \in H^{(2,1)}(X)$, $G^* \in H^{(1,2)}(X)$,
and $\tilde{G}^*$ vanishes.  Therefore, 
we can see directly from the form of the auxiliary fields
that an expectation value for $G$ lying entirely in
$H^{(2,1)}(X)$ preserves the same supersymmetry as
a D3-brane placed in the CY background, consistently with
\refs{\GranaJJ,\GubserVG,\GranaXN,\GiddingsYU}.

\subsec{Massless hypermultiplets}

In 
type IIB string theory, 
the K\"ahler moduli of the Calabi-Yau 
live in massless hypermultiplets.
For a given element $\omega^a$ of $H^{(1,1)}(X)$,
the four scalars correspond to the metric perturbation
$g_{i\jb} = g\omega^a_{i\jb}$, the NS-NS two-form potential
$b_{i\jb} = b \omega^a_{i\jb}$, the RR two-form potential
$c_{i\jb} = c \omega^a_{i\jb}$, and a scalar which is the 
4d dual of the RR 4-form potential $C^{(4)}_{\mu\nu i\jb} = 
c_{\mu\nu}\omega^a_{i\jb}$.  These should form a triplet
and a singlet under the $SU(2)$ R-symmetry.

Ideally we would embark on a discussion of the auxiliary fields
for IIB hypermultiplets.  However, as we will see,
these fields, as well as the NS-NS auxiliary fields
in IIA vector multiplets, correspond to ``mirrors of NS flux''
\refs{\VafaWI,\KachruSK,\NSmirror}, and are not yet well-understood.
Therefore we will confine our discussion of hypermultiplets 
to type IIA string theory compactified on CY threefolds,
leaving the IIB discussion for future work.
Assuming a suitable generalization 
of mirror symmetry can be formulated, 
the structure of auxiliary fields for hypers 
which we find in IIA will also govern the IIB physics.

In 
the hypermultiplets of
type IIA string theory, 
the complex structure deformations again play a starring role.
There
are four real scalars in this multiplet. Two derive from
the complex scalar corresponding to deformations of the
complex structure.  The other two derive from the
RR four-form field strength $F^{(4)}$.  If $\omega^m$
is a basis element of $H^{(2,1)}$ corresponding to
a complex structure deformation, then 
$b^{21}$ complex vector field strengths $F_{\mu}^m$ arise via:
\eqn\vffromfourform{
   F^{(4)} = \sum_m F_{\mu}^m \omega^m_{\bi j k}
   dx^{\mu} \wedge d\bar{z}^{\bi}\wedge dz^j \wedge dz^k
+{\rm h.c.}~.
}
The Bianchi identities for $F^{(4)}$ and the fact that
$\omega^M$ is a closed form imply that 
$F$ is the derivative of a scalar.
\eqn\rrscalar{
   F_\mu^m = \p_\mu \phi^m
}
Thus we have two complex scalar fields for each $(2,1)$-form.

We can write a scalar superfield $\CH$ for the hypermultiplet
which satisfies the ``twisted chiral'' constraints:
%
\eqn\twistedchiralconstraints{
\eqalign{
   \bar{\nabla}_{\dot{\alpha}} \CH & \equiv 
      \left(- \frac{\p}{\p\bar{\theta}^{\dot{\alpha}}} 
      - i \sigma^\mu_{\beta \dot{\alpha}}
      \theta^{\beta} \p_{\mu}\right) \CH = 0 \cr
   \hat{\nabla}_{\alpha} \CH & \equiv \left( 
      \frac{\p}{\p\hat{\theta}^{\alpha}} 
      + i \sigma^\mu_{\alpha \dot{\beta}}
      \hat{\bar{\theta}}^{\dot{\beta}} \p_{\mu}\right) \CH = 0\ .
}}
The superspace expansion for the
massless hypermultiplets is \berksieg:
\eqn\twohyper{
\eqalign{
   \CH^a & = w^a + \theta^\alpha \chi_\alpha^a
   + \bhthet^{\dot{\beta}} \hat{\bar{\chi}}_{\dot{\beta}}^a
   + \theta^2 y^a + \hthet^2 \hat{\bar{y}}^a \cr
   \ \ & \ \ + \theta^\alpha \bhthet^{\dot{\beta}}
   \sigma^\mu_{\alpha\dot{\beta}} F_\mu^a \cr
   \ \ & \ \ + \theta^\alpha \bhthet^2
   \eta^a_{\alpha}
   + \bhthet^{\dot{\beta}}\theta^2
   \hat{\bar{\eta}}^a_{\dot{\beta}} \cr
   \ \ & \ \ + \theta^2 \bhthet^2 C^a
}}
Here $w, F_\mu, C$ are all complex.  On-shell, $w$ can
be identified with a deformation of the complex structure
and $F_\mu = \p_\mu \phi$ is the corresponding RR axion
scalar field strength.

All of the worldsheet arguments given for the type IIB
vector multiplets apply here in identifying $y,\hat{\bar{y}}$.
While $\hat{\bar{y}}$ is related to $w$ by $\hat{\bar{Q}}$
in type IIA, the worldsheet current $\hat{\bar{Q}}(\hat{z})$
for the spacetime supercharge depends on the same spectral
flow operator, with $U(1)_R$ charge $3/2$, as the 
current $\hat{Q}$ does in type IIB string theory.
Therefore, if we use ``CFT coordinates''
we can identify:
\eqn\cfthyperauxiliary{
\eqalign{
   y^m & = \tau^m + h^m \cr
   \hat{\bar{y}}^m & = \tau^m - h^m\ ,
}}
where $\tau^m,h^m$ are given in \fluxescft.
Similarly, if we choose the periods of the $(3,0)$ form
as our moduli space coordinates, we find that:
\eqn\periodhyperauxiliary{
\eqalign{
   y^a & = \int_{A^a} \left(\tilde{T} + \tilde{H}\right) \cr
   \hat{\bar{y}}^a & = \int_{A^a} \left(\tilde{T} - \tilde{H}\right)\ .
}}
and if we choose the dual periods:
\eqn\dualperiodhyperauxiliary{
\eqalign{
   y_a & = \int_{B_a} \left(\tilde{T}+\tilde{H}\right) \cr
   \hat{\bar{y}}_a & = \int_{B_a} \left(\tilde{T} - \tilde{H}\right)\ .
}}
Note that the mixed $\theta \hat \theta$ term 
in \twohyper\ is a propagating RR axion. 
The corresponding vertex operator for this field is:
\eqn\rraxionvo{
V^{(-\half,-\half)}_{F^m_\mu} = 
e^{- {\phi\over 2}- {\hat \phi \over 2}}
F_\mu^m 
S^{\alpha}\sigma^\mu_{\alpha \dot{\beta}}\hat{\bar{S}}^{\dot{\beta}}
S^{-,\dot{a}} (C\Gamma)^{ij\bar k}_{\dot{a}\dot{b}}
\hat{S}^{-,\dot{b}}\omega^m_{ij\bar{k}}\ .
}
Again, $S^\alpha,\hat{\bar{S}}^{\dot{\beta}}$ are spin fields
describing 4d spinors, while $S^{-,\dot{a}}, \hat{S^{-,\dot{b}}}$
are spin fields for internal $6d$ spinors.

A final check of these identifications is the solution in
\refs{\MaldacenaYY}\ corresponding to NS5-branes wrapped
on a holomorphic 2-cycle.  This is a solution
in both type IIA and IIB string theory, as it
contains vevs for NS fields only, and it breaks
either $Q$ or $\hat{Q}$, depending on the relative sign
of $H$ and $T$.

\subsec{Comparison to the hybrid formalism}

Berkovits and Siegel \refs{\BerkovitsWR,\berksieg}\ 
have constructed a manifestly spacetime supersymmetric
worldsheet theory for type II and heterotic strings
compactified on a Calabi-Yau background.  In this
formalism, the superspace coordinates $x,\theta,\hat{\theta}$
appear as worldsheet fields, and together with an additional
boson $\rho$ describe a CFT with a nonlinearly realized
$\CN=2$ superconformal field theory.  This
SCFT is then combined with a (twisted) $c=9$ $\CN=2$ 
superconformal field theory which
is the usual CFT describing the Calabi-Yau
compactification.  The $\CN=2$ superconformal is gauged.
There is no barrier to writing 
down the worldsheet theory in the presence of nonvanishing RR fields.

This appears to be an ideal formalism for the models discussed
in this paper.  Nonetheless, there remain some
things to be understood.  In \refs{\BerkovitsWR,\berksieg},
the authors argue that the $\CN=2$ physical state constraints
require that $D_{ij}$ and $y$ vanish.  In view of the
results of \refs{\geometrictransitions-\CachazoJY},
and our identifications of the auxiliary fields,
there appears to be a problem.  Presumably the physical
state constraints are modified in the presence
of nonvanishing flux.~\foot{We would like
to thank N. Berkovits for correspondence on this issue, 
and for suggesting this resolution.}
We will leave this question for future work.

\newsec{Engineering soft breaking terms in string theory}

Now that we have identified the auxiliary fields for the
closed string moduli, we can ask how these fields
appear in the low-energy effective action as
coefficients of SUSY-breaking operators.  We will
start by reviewing the argument in \refs{\VafaWI}\ for
the superpotential in \refs{\GVW,\TaylorII}, and
then discuss fluxes which break $\CN=1$ supersymmetry completely.

\subsec{The GVW superpotential}

Given the representation of the vector multiplets 
in type IIB string theory in terms of $\CN=2$ chiral 
superfields $V^a$, we can write the low energy Lagrangian for
these fields locally in the vector multiplet moduli
space in terms of a holomorphic prepotential $\CF(t)$,
such that the dual variables $F_a$ can be written as
$F_a = \p_a \CF$.  The low-energy effective Lagrangian can be written
in terms of an unconstrained 
$V$ and a chiral superfield $V_D$ which acts as a 
Lagrange multiplier \refs{\AntoniadisVB,\PartoucheYP}:
\eqn\prepotlagrang{
   \CL = \int d^2 \theta d^2 \hat{\theta}
   \left( \CF(V) - \sum_a V_{D,a} V^a \right)\ .
}
The equation of motion for $V$ is
$$ V_{D,a}(V) = \p_a \CF\ . $$
Integration over $V_D$ implements the constraint \supercon\ on $V$.

Vafa \refs{\VafaWI}\ has shown in a type IIA model that
when the auxiliary fields in $V, V_D$ have nonvanishing
vevs, one can expand \prepotlagrang\ in these auxiliary fields
and get the superpotential described in \refs{\GVW,\TaylorII}.
For a proper choice of vev, the vevs of these
auxiliary fields break $\CN=2$ supersymmetry
to $\CN=1$, as we have discussed in the previous section.
For example, we can choose a flux $G \in H^{(2,1)}$,
which leaves unbroken the supersymmetry that generates
superspace translations along $\theta - i \hat{\theta}$.
Indeed, upon integrating over $\theta + i \hat {\theta}$
in \prepotlagrang, one finds a $\CN=1$ superpotential term
which is linear in the auxiliary fields, and therefore the
fluxes:
\eqn\expansion{
   W = - D_D \cdot \CA (x, \theta - i \hat{\theta})
}
where $D_D = \int_B G$ is the auxiliary field
multiplying $(\theta + i \hat{\theta})^2$, and $\CA$ is the
$\CN=1$ chiral superfield one gets from 
translating $V$ in the $(\theta - i \hat \theta)$ direction 
of superspace:
%
\eqn\chiralexp{
   \CA = w + (\theta - i \hat{\theta}) \zeta 
   + (\theta - i \hat{\theta})^2 D\ .
}

It is known \refs{\GVW,\TaylorII}\ from other arguments 
that nonvanishing $G$ induces a superpotential
for complex structure moduli:
\eqn\GVWsuperpot{
   W(t) = \int G \wedge \Omega(t) = \sum_a
   \int_{A^a} G \int_{B_a} \Omega - \int_{B_a} G \int_{A^a} \Omega \ .
}
\expansion\ has the right form, but it is missing the
piece proportional to $\int_{A} G \int_B \Omega = F_a \int_A G$.
This is as it should be.  A spurion superfield, as used in
\refs{\ggsoft}, should be nondynamical: for these superfields,
we can tune the vevs of the component fields by hand
without going off shell.  The auxiliary fields $D_{ab}$ for
$V$ lie in supermultiplets for propagating particles.
Vevs for these fields must arise from spontaneous
SUSY-breaking, which arise in global $\CN=2$ theories via
Fayet-Iliopoulos terms 
\refs{\AntoniadisVB,\AntoniadisKI,\PartoucheYP}.



We can give an explicit example where the superfields
for the dual periods
can be spurions, allowing a desciption of explicit soft SUSY breaking
to $\CN=1$.
Following \refs{\CachazoJY}, choose the
Calabi-Yau hypersurface
\eqn\civlocalmodel{
   y^2 + u^2 + v^2 = W'(x)^2 + f(x)
}
where $W(x)$ is an $(n+1)$st-degree polynomial
(so that $W'(x)$ is an $n$th-degree polynomial)
and $f(x)$ is a degree $(n-1)$ polynomial.
The complex structure moduli space has been described
in \eg\ \refs{\CachazoJY}.
There are $n$ complex structure deformations
which are normalizable, in the sense of having a
finite kinetic term, and which can be described
as deformations of $f$.  These control the volumes
of $n$ compact, independent, nonintersecting
cycles we will label $A_a$,
localized near the zeros of 
$W^\prime = \prod_{a=1}^n ( x - x_a) $.
There is a set of noncompact dual cycles $B^a$
which intersect $A_a$ once and extend to infinity.

The periods $t^a$ along $A_a$
are propagating vector multiplets, while
the dual periods for $B^a$ can be treated as spurions:
they  are clearly not normalizable, and
the fluxes through these noncompact cycles are
not quantized and may be tuned continuously from zero.
In this case, the superpotential \expansion, which is the
realization of \GVWsuperpot\ in this case,
is linear in the periods of
the A-cycles.  The corresponding scalar potential
will only vanish when the inverse metric on the
moduli space vanishes, which occurs when the
A-cycles shrink to zero volume \refs{\PolchinskiSM}.

For flux through compact cycles,
we can explicitly identify fields which
control the FI parameters.
For illustration, focus on a single pair of compact
$A$ and $B$ cycles. 
A D3-brane on a special Lagrangian representative of 
the homology class of the B-cycle is a 
hypermultiplet which is 
{\it magnetically} charged under the 
vectormultiplet built on $t$.  
Its mass is determined by the volume of the 
B-cycle $ m_Q = F_t = \int_B \Omega $.
Call the scalars in this multiplet $Q_i, i = \pm$.  
$\CN=2$ supersymmetry implies 
a superpotential coupling 
$ W_{\CN=2} = Q_+ F_t Q_- $
which encodes the mass.
Further, the magnetic gauge field 
(the gauge field associated to the B-cycle) 
is related by $\CN=2$ SUSY to a triplet of auxiliary fields $D_{Dij}$, 
and the electric charge of $Q$ under this gauge field 
implies a coupling to the scalars of the form
$$ Q_i Q_j D_{Dij} .$$
Therefore, a vev 
$$ \vev{ Q_i Q_j} = r_{ij} $$ 
acts precisely as the magnetic FI coupling for this 
vectormultiplet $r D_D$.  
Plugging this into the $\CN=2$ Lagrangian
and varying $D_D$, one learns that the vev of the 
auxiliary field in the electric vectormultiplet is 
$$ D_{ij} = r_{ij} ~.$$
We have identified $D$ with the flux through 
the A-cycle.  It will be interesting to understand
the origin of flux quantization from this
perspective.

\subsubsection{Effect of fluxes on wrapped branes which are 
particles}

Before continuing, we would like to discuss how
such fluxes affect the physics when D-branes wrapping vanishing
cycles enhance the closed-string gauge group to a nonabelian
gauge group.  In this case the wrapped branes
are charged under the vector multiplets controlling
the periods of the cycles that are wrapped.  Branes
wrapping dual cycles are magnetically charged. 
However, the background fluxes do not allow
charged states made out of these wrapped branes.
If we make a particle state by wrapping a Dp-brane 
on a p-cycle $W$ 
which has p-form RR flux through it,
then because of the worldvolume coupling
$$ \int _{Dp} F \wedge C_{p-1} = - \int_{Dp} A \wedge F_{p} ,$$
where $F = dA $ is the worldvolume gauge field,
the Gauss' law on the Dp-brane is modified to
\eqn\guasslaw{
0 = {\delta S\over \delta A_0} = \int_{W} F_{p} + \delta( \del F1)  .}
The brane can't solve its own Gauss' law
unless it has the right number 
$q = \int_W F_p $
of F-strings ending on it 
\refs{\PolchinskiSM,\baryonsandbranesinads}.
Therefore, such branes must come in pairs 
comprised of a brane and an anti-brane, 
with $q$ F-strings stretching between them.

Furthermore, the flux through the 3-cycle
will try to prevent that cycle from shrinking.
This is clear on general energetic grounds, and
is borne out by 
extremizing the
GVW superpotential.
Both of these effects are in harmony
with what is known about the low energy effective
action in the presence of such a flux.  As pointed
out in \refs{\TaylorII} in the context of type IIA
string theory, such a flux corresponds
precisely to breaking $\CN=2$ SUSY to $\CN=1$ by
giving a mass to the scalar in the closed-string 
vector multiplet,
while branes wrapping the dual cycles will
be light magnetic monopoles, with a mass controlled by the dual period.
Thus, the terms in \refs{\TaylorII}\ that arise from the flux
are known to lead to monopole condensation and
confinement \refs{\SeibergRS,\GreeneDH}.

\subsec{Soft breaking parameters through F-terms}

Gauge theories with superpotentials for
adjoint scalars are easy to realize in 
type IIB string theory \refs{\bdlr,\kklma}.
As an example, begin with D5-branes wrapping
holomorphic 2-cycles in a Calabi-Yau manifold. 
There are adjoint chiral superfields arising from holomorphic
deformations of the supersymmetric cycle
and of open string gauge bundles living on those
cycles.  To all orders in perturbation theory,
the superpotential for these fields is determined
entirely by the obstructions to finite holomorphic
deformations of these cycles, and so
the superpotential couplings depend entirely
on complex structure moduli \refs{\bdlr,\DouglasVM,\DouglasGI}.
In other words, the superpotential can be written 
as
\eqn\gensuper{
   W = W(t_a, \Phi)
}
where $t_a$ are the complex structure moduli.
At energies low enough that gravity decouples from the
D-brane gauge theory, $t_a$ may be treated as couplings.
Supersymmetry is broken if these couplings are treated
as chiral superfields and the auxiliary components
of these superfields are given vacuum expectation values.

For example, for the quadratic and cubic terms in the superpotential,
\eqn\relsuper{
   W_{rel} = \half t_2 \tr \Phi^2 + \frac{t_3}{3!} \tr \Phi^3
}
the SUSY-violating terms are soft and do not spoil
the ultraviolet properties of the theory.  If we promote
$t_{2,3}$ to superfields $T_{2,3}$, the most general expectation
value preserving Lorentz symmetry is:
\eqn\csshift{
   T_i = t_i + \theta^2 F_i\ .
}
Following the discussion in \S2.1, $F_i$ will
correspond to a vev for the flux $\bar{G}\in H^{(2,1)}(X)$,
or equivalently a vev for $G\in H^{(1,2)}(X)$.  This
breaks the spacetime SUSY unbroken by the D5-brane (at large
volume).
  
The action which derives from \relsuper\ is:
\eqn\ftermshoft{
  \int d^4x d^2 \theta~ W(T,\Phi) + {\rm h.c.}\ =
  \int d^4 x d^2 \theta~ W(t,\Phi) + {\rm h.c.} +
  S_{sb}
}
where the SUSY-breaking terms are:
\eqn\softaction{
   S_{sb} = \int d^4 x \left(
      F_2 \tr \phi^2 + F_3 \tr \phi^3 + {\rm h.c.}\right)\ .
}
The SUSY-breaking part of the action includes only
the scalar components of $\Phi$.  $\phi$ has two real components
and the mass term $F_2$ induces a mass matrix with positive
and negative mass squared.  $F_3$ induces a SUSY-breaking Yukawa coupling.

We can give a partial worldvolume argument for these SUSY-breaking terms.
A D5-brane couples minimally to the RR 6-form
potential, $ S_{D5} \ni \int_{D5} C^{(6)}$.
We are interested in a D5-brane which 
wraps a curve in the CY  and fills the transverse $R^4$.
With this in mind,  decompose the six-form potential
as $C^{(6)} = dv \wedge c^{(2)}
+ \dots $
where 
$ dv = dx^0\wedge dx^1 \wedge dx^2 \wedge dx^3$ 
and
$\dots$ indicates other polarizations which will 
not concern us.
The expression of minimal coupling above
is shorthand for the 4d Lagrangian density 
\eqn\minimalcoupling{ L_s =  \int_{\CC} \tr \Phi^\star c^{(2)}\ ;}
following \WittenEP, 
we have described the part of the worldvolume of the brane in the CY 
as an embedding from an abstract curve $\CC$ 
into the CY. In this description, 
the RR potential must be pulled back to the branes 
by the embedding fields $\Phi$,
and the trace indicates that this pulled-back potential 
is really a matrix fuction
of the matricized embedding coordinates.
Next, extend the map $\Phi$ to 
a three-chain $\Xi$ with
%
which bounds the curve $\CC$:
$ \del \Xi = \CC$.
The coupling \minimalcoupling\ 
can be rewritten as an integral over the bounding 3-chain, as 
\eqn\intermsofg{ L_s = \int_{\Xi} \Phi^\star g^{(3)}\ ,}
where $g = dc$.  By self-duality
of RR fluxes in type IIB, 
the seven-form flux satisfies 
\eqn\selfdual{F^{(7)} = dv^{(4)} \wedge g^{(3)} 
= \star F^{(3)} \ .
}
Therefore, $ g^{(3)} = \star_{CY} F^{(3)}$, and the
induced bosonic potential can be written as:
\eqn\whatwewant{ L_s=  \int _{\Xi} \Phi^\star 
\left( \star_{CY} F^{(3)} \right)\ .}
In the presence of a constant RR axion $C^{(0)}$,
the Witten effect modifies the minimal coupling to
$S_{D5} \ni \int_{D5} \left( C^{(6)} - C^{(0)} B^{(6)} \right) $
where $d B^{(6)} = \star H$ is the field to which 
NS5-branes couple electrically.
Retracing our steps, \whatwewant\ becomes
$ \int_{\Xi} \Phi^\star 
\left( \star_{CY} \left( F - C^{(0)} H \right) \right)$.
Let us consider for the moment $C^{(0)} = 0$.
Let us turn on the RR 3-form potential such that
$F = F^{(2,1)} + F^{(1,2)}$.  The terms on the right hand side
are imaginary self dual and imaginary anti-delf-dual, respectively.
Therefore,
\eqn\rrdualflux{
\star F = i (F^{(2,1)} - F^{(1,2)})\ ,
}
and the potential is:
\eqn\evenbetter{
 \int_\Xi \Phi^\star 
\left( i F^{(2,1)} - i F^{(1,2)} \right) ~.}

Compare this coupling with Witten's description 
\WittenEP\ of the
obstruction {\it super}potential 
as a function of the embedding fields
$\Phi$ and the holomorphic threeform $\Omega(t)$:
\eqn\wittenobstruction{
W(t,\Phi) \propto \int_{\Xi} \tr \Phi^\star \Omega(t) .
}
Now let $\eta = \theta + \hat{\theta}, \xi = \theta - \hat{\theta}$.
$\eta$ corresponds to the supersymmetry which is preserved by the
D5-branes.  We may rewrite \threeformsuperfield\ as:
\eqn\fivebranethreeform{
\Omega(\eta,\xi) = \Omega(t) + \eta^2 (\tilde{F} + \tilde{T}) + \xi^2
(\tilde{F} - \tilde{T})
+ \ldots
}
Turning on $F$ without turning on $T$ clearly breaks the remaining
SUSY preserved by the D5-brane.
Integrating \wittenobstruction\ along the superspace directions 
preserved by a D5-brane, 
using the identifications of \S2, 
one indeed finds that the potential \evenbetter
matches the SUSY-breaking potential derived from \wittenobstruction.


\subsubsection{An example}

The canonical example of a curve with 
obstructed deformations is found in 
the following hypersurface singularity in $\IC^3$:
\eqn\localCY{
u^2 + v^2 + y^2 = x^{2n}~.
}
A small resolution of this singularity introduces a
homologically nontrivial $\IP^1$ at the origin
of $\IC^3$.  The normal bundle of this
curve has a holomorphic section with
an $n$th order obstruction \refs{\reidca}.
N D5-branes wrapped around such a cycle
have as their low-energy degrees of freedom
a $U(N)$ gauge field and an adjoint scalar
$\Phi$ with superpotential $W = \Phi^{n+1}$ 
\refs{\bdlr,\kklma,\CachazoJY}.

Let us choose $n=2$ to get a cubic superpotential.
A mass term is turned on if one deforms
\localCY\ so as to split the singularity:
\eqn\localsplit{
u^2 + v^2 + y^2 = ( x^2 - a^2)^2 
}
The geometry has a compact $S^3$ which can
be described as follows.

For each $x$, \localsplit\ describes an $A_1$ ALE space.
This space has a resolved $S^2$ with radius
$r = |x^2 - a^2|$, described by a real slice of
\localsplit\ at fixed $x$. If we fiber this $S^2$
over the line between $x=a$ and $x=-a$, we sweep out
an $S^3$ which we will denote as $C$ \refs{\CachazoPR}.  
Following the discussions in \refs{\CachazoJY,\dv},
there are two noncompact ``B-cycles'' which are
the 2-spheres in the $(y,u,v)$ directions
fibered over lines running from $x=\pm a$ to infinity,
and which we will label $B_{\pm}$.
 
We can resolve the conifold singularities
at $x=\pm a$ with normalizable deformations
of \localsplit\ as follows:
\eqn\localsplitdef{
   u^2 + v^2 + y^2 = (x^2 - a^2)^2 + bx + c 
}
where $b,c$ are the normalizable deformations.
These split the double points at $x = \pm a$, introducing
finite size three-cycles $A^\pm$ which are dual to $B^\pm$ and
which intersect $C$ with intersection number $\pm 1$.

On the other hand, if we perform a small resolution,
we introduce one $\IP^1$ at each of $x = \pm a$.
These are homologous.  $C$ becomes a three-chain
whose boundary is the difference 
of these two 
$\IP^1$s, and $B^\pm$ become three-chains as well.  The
parameter $a$ is a nonnormalizable complex structure
parameter which controls the distance between the
two rational curves at $x = \pm a$.

If we perform a such a small resolution and
wrap a D5-brane around one of the $\IP^1$s, there will
be a single scalar field parameterizing
holomorphic deformations of the $\IP^1$,
and the superpotential will take the form
\refs{\bdlr,\kklma}:
\eqn\splitsuper{
   W = \tr \left(\frac{1}{3}\Phi^3 - a^2 \Phi \right)\ .
}
By changing variables to $\tPhi = \Phi + a$, and expanding
$W$ mear $\tphi = 0$, one can see that
$a$ is a mass parameter.
The SUSY-breaking flux that $F_a$ corresponds to
will be $\bar{G}$ with support
on the three-cycles whose periods depend on $a$.
Such fluxes are 
three-forms which do not die off at infinity.  
We can understand better
how these
can arise when we embed \localsplit\ in a more 
complete model.

\subsubsection{Three-cycles for nonnormalizable deformations}

Let us explain 
how \localsplit\ describes a patch of a slightly more global  
description of a compact geometry.
Let $Y$ be a genus-g curve $\CS_g$ of singular $A_1$ ALE fibers embedded
in some Calabi-Yau \refs{\CandelasDM,\KatzHT}.
A particular example \refs{\CandelasDM}
is the genus-three curve of $A_1$ singularities
at $z_1 = z_2 = 0$ in the hypersurface
\eqn\newhyper{
   z_1^8 + z_2^8 + z_3^4 + z_4^4 + z_5^4 + \ldots
}
in $\IC\IP_{1,1,2,2,2}$.  We can now blow up the
orbifold singularity, giving us a family $S_g$ of
$\IP^1$s.  D5-branes wrapped around an
$\IP^1$ fiber have a moduli space which is this
Riemann surface \refs{\KachruAN}.

However, there is a set of non-toric (in this realization)
complex structure deformations
which can be related to harmonic forms $\omega\in H^1(S_g)$.
Turning on such deformations leave in general
$(2g -2)$ rational curves at points on $S_g$ corresponding
to zeros of $\omega$.  Now, we can describe 
a star-shaped patch of this Riemann surface 
by a complex coordinate $x$.
Locally near some set of $n$ zeros of $\omega(x)$,
this can be modeled precisely by \civlocalmodel\ with $f(x) = 0$,
and D5-branes have a superpotential described in $W'(x)$,
as can be inferred from \refs{\KachruAN}.  
The nonnormalizable deformations in \civlocalmodel\
are in fact the ``non-toric'' deformations of $Y$.
Furthermore,
one may blow down the $A_1$ fibers and pass through
an extremal transition.  In the local model this
can be described via a deformation $f(x)$ in \civlocalmodel.

A natural set of three-cycles $E^i$ corresponding to the
non-toric deformations
are fibrations of $S^2$s in the $A^1$ fiber over the
cycles of the Riemann surface $\CS_g$ \refs{\CandelasDM, \KatzHT}.
From the point of view of the local patch these
three-cycles exist because of nontrivial structure 
hidden at infinity.
In this example, the periods of $C$, and $B^\pm$ 
and some of $E^i$ all depend on $a$.

\subsec{Gaugino mass terms}

The kinetic term for an $\CN=1$ vector 
can be written as a half-superspace integral
\eqn\worldvolumekinetic{
\CL = \int d^2 \theta~ \tau_{YM}( \phi) W_{\alpha}W^\alpha
}
where $\tau_{YM}(\phi)$ is the chiral superfield 
gauge coupling function.  
From \worldvolumekinetic\ we see that 
a mass for this gaugino results
from giving vev to the $\theta^2$ component 
of the chiral multiplets on which $\tau_{YM}$ depends:
\eqn\gauginomass{
   m_{{\rm gaugino}} = F^a \p_a \tau_{YM}
}
where $F^a$ is the auxiliary field corresponding to $\phi$.

Which closed string fields appear in $\tau_{YM}$ depends on
how the gauge symmetry is realized.  For space-filling
B-type D-branes in type IIB, the gauge coupling is 
controlled, to leading order, 
by the volume of the cycle the brane
wraps, which is in turn controlled by the K\"ahler moduli.
These live in hypermultiplets.  Therefore, the gaugino masses
will depend on the auxiliary fields $y,\hat{y}$.  We understand
these fields somewhat better in type IIA models.  The mirror
D-brane configuration is a D6-brane wrapped on a special
Lagrangian 3-cycle $A\subset X$ 
with phase $e^{i\gamma}$.  The gauge coupling is a function
of the volume of this cycle, and is in fact linear
in this volume to leading order.  
The tree level gauge coupling function is 
%
\eqn\gaugecouplingIIA{
    \tau_{YM} = { \theta \over 2\pi} 
+ i \frac{4\pi}{g_{YM}^2} = 
{i\over g_s} {\rm Re}~ e^{i\gamma} \int_A \Omega
+ \int_A C 
}
The gaugino masses will arise from the auxiliary field
for $t$.  For example, if we break the SUSY charge $\hat{Q}$,
the gaugino mass is proportional to
\eqn\gauginomass{
    m_{\rm gaugino} = {\rm Re}~ e^{ i \gamma } y 
}
where $y = \int_A (H - T)$
is the flux through the cycle on which the
D6-brane is wrapped.

Another possible source of gauge dynamics is from
vanishing cycles, if there are branes
which can wrap these cycles and which live at a point in
the four-dimensional spacetime.  We will focus on
singularities arising from D3-branes wrapped
around vanishing 3-cycles in type IIB string theory.
If these branes have spin-1 states, they lie in vector multiplets which
become massless when the geometry becomes singular.
Away from the singularity, the gauge symmetry is 
completely broken down to the maximal torus of the original
group.  This maximal torus consists of the vectors
in the vector multiplets that control the volume of the vanishing
cycles, which are perturbative degrees of freedom in type IIB string
theory.  The dynamics of these vector multiplets are controlled
by the perturbative prepotential.

The difference from the previous example is that
these gauge symmetries arise already in the
$\CN=2$ theory before SUSY was broken.  Therefore,
the gauge dynamics is a function of the
vector multiplets, in contrast with the wrapped-brane example
above.

The gauge coupling of the $\CN=2$ theory is the second derivative
of the prepotential:
\eqn\pertgaugecoupling{
   \tau_{ab} = \CF_{ab} (V)
}
Choose for example a single abelian vector multiplet $V$
with a scalar $t$ equal to the
period of a 3-cycle $A$.  A D3-brane wrapped
on this three-cycle is charged under the vector
multiplet.  If it has a spin-one excitation, the
perturbative $U(1)$ gauge theory is enhanced to $SU(2)$.
The gauge coupling is
$$ \tau_{YM} = \CF_{aa}(V) \ . $$
If we break the supersymmetry corresponding to $\hat{Q}$, $t^a$
lives in a chiral multiplet with respect to $Q$ that includes
$t^a$ and $y^a$, where we replaced $w^a$ with $t^a$ in \twohyper.
Hence $D_{--}^a = \int_A (H-T)$ is the gaugino mass.

\subsec{Scalar mass terms}

Another phenomenologically important SUSY-breaking
mass term is of the type:
\eqn\squarkmass{
   S_{sb} = \int d^4 x M \phi^{\dagger} \phi
}
which gives a positive mass to both the
real and imaginary parts of $\phi$.  In particular
squark masses are of this form in the minimal
supersymmetric standard model.

In general such masses arise can arise via spurions
in two ways, depending on whether the spurion is a chiral or
a vector multiplet.  Imagine
a K\"ahler potential for some scalar field $\phi$,
as a function of moduli coordinates $t$.  
If the spurion is a chiral multiplet,
let the corresponding superfields be $T, \Phi$.
If $T$ has as an auxiliary field $F$,
a term in the K\"ahler potential of the form
\eqn\kahlerforsquarks{
   K(t,\phi) = \frac{1}{\Lambda^2} T^{\dagger} T \Phi^{\dagger}\Phi\
}
leads to a mass term for $\phi$:
\eqn\examplemass{
   S_{sb} = \int d^4 x \frac{|F|^2}{\Lambda^2} |\phi|^2\ .
}
For general open string fields, the Kahler
potential can be a function of both vector
and hypermultiplets, in which case fluxes
can generically lead to squark masses.\foot{The form 
of these couplings which arises 
in a supergravity approximation are determined  
in \refs{\CamaraKU, \Grananew}.}

Similarly, we can imagine that the scalar field $\phi$ is
charged under a nondynamical vector multiplet, so that
the action is
\eqn\kahlerforsquarksII{
   S = \int d^4 x d^4 \theta ~\Phi^{\dagger} e^V \Phi 
}
where $V$ is a vector superfield.  A vev for the auxiliary scalar $D$
in this vector multiplet will induce a soft scalar mass $D|\phi|^2$.
However, in general open strings will not be charged under closed
string gauge groups.  Such a term might arise if the squarks
are charged under a weakly coupled $U(1)$ propagating
on a brane distinct from the branes carrying the standard
model gauge symmetries.  If SUSY is broken on this other
brane, such a coupling may emerge.

\newsec{The Dijkgraaf-Vafa conjecture and softly broken SUSY}

Dijkgraaf and Vafa \dv\ have
argued that the full set of F-terms in the low-energy
effective action of $d=4$, $\CN=1$
SUSY gauge theories can be computed using
the saddle point approximation to
an auxiliary matrix integral.  
The original motivation for this conjecture arose from
the appearance of such field theories as the low-energy
description of open strings ending on D-branes, in type II
string theory compactified on a Calabi-Yau threefold,
precisely the class of theories we have discussed above.
Since then, Dijkgraaf \etal\ have provided
a perturbative derivation of this conjecture directly in the
quantum field theory \refs{\dvpert,\cdswone}, and Cachazo
\etal\ \refs{\cdswone}\ have given a nonperturbative argument via the
Konishi anomaly.


These arguments might appear to fail when supersymmetry
is broken dynamically or via explicit soft SUSY-breaking terms.
Important elements of the proof outlined in \refs{\cdswone}\ depend
crucially on the fact that the SUSY charges annihilate the
vacuum when SUSY is unbroken: 
the statements that correlators of gauge-invariant chiral fields
are constant \refs{\nsvza,\rv,\amatietal,\cdswone},
that they are holomorphic
in the superpotential couplings, 
and that the vacuum expectation
values of the non-chiral terms in the Konishi anomaly
equation vanish.


We find that the Dijkgraaf-Vafa proposal 
for computing low-energy superpotentials 
retains some force 
when supersymmetry is broken
via a class of explicit soft supersymmetry-breaking terms,
if these terms are small compared to the other
dynamical scales of the theory.~\foot{
The paper \refs{\brandetal}\
uses the techniques in \refs{\cdswone}\
to examine the IYIT model 
\refs{\izya} of dynamical supersymmetry breaking.}
This follows earlier discussions of 
perturbative \refs{\ggsoft}\ and
nonperturbative \refs{\EvansIA}\ nonrenormalization
theorems for F-terms when supersymmetry is softly broken. 
The essential point is that so long as one
is asking holomorphic questions, one will
get holomorphic answers.  
For this to be possible, 
one must be able to isolate the 
holomorphic parts of 
amplitudes.
We may do this
as long as the SUSY-breaking terms are genuinely soft,
meaning that they are indeed relevant operators
whose effects are negligible in the UV.
In such a circumstance,
amplitudes will be analytic in the soft-breaking 
couplings, 
%
and one can isolate
the holomorphic part 
by expanding in the antiholomorphic
couplings and keeping the constant term.

Although one may compute the F-terms in these examples, one
must face the fact that they no longer control
the correlators of chiral operators; D-terms infect all of the
answers and the effects are calculable only when the
soft breaking terms are small.  Furthermore, once the soft SUSY-breaking
masses are larger than the dynamical scales of the theory,
the results we extract using 
supersymmetry fail to be a useful guide to low-energy physics.

In the remainder of this section we will make these statements
more precise.  After quickly reviewing the
Dijkgraaf-Vafa conjecture in the simplest case
without SUSY breaking ($U(N)$ gauge theory with an
adjoint chiral scalar), 
we will demonstrate first the sense 
in which the perturbative proofs
in \refs{\dvpert,\cdswone}\ remain valid
when explicit soft SUSY breaking terms are
added to the action, and then how 
the nonperturbative
arguments of \refs{\cdswone}\ remain valid.\foot{
See \refs{\konvez}\ for an 
early discussion of the low-energy effective action
for a system with dynamical SUSY breaking.}

\subsec{A brief review of the Dijkgraaf-Vafa conjecture}

To understand the conjecture let us describe 
the simplest case, that of an $\CN=1$ $U(N)$ gauge
theory in four dimensions, coupled to an adjoint chiral
superfield $\Phi$. The gauge coupling and theta angle
can be written as a complex coupling 
$\tau = \frac{4\pi i}{g^2} + \frac{\theta}{2\pi}$.
Let $\Phi$ have a tree-level superpotential
$W_0(\Phi)$; if the superpotential has $k$ extrema
and one chooses the vev of $\Phi$ such that
$N_i$ eigenvalues reside in the $i$th critical point,
the gauge group is broken to:
\eqn\broken{
  U(N) \longrightarrow
  U(N_1)\times U(N_1)\times \cdots \times U(N_k).
}
This theory confines, and 
the low-energy degrees of freedom are believed 
to be described by the ``glueball'' superfields
\eqn\glueballvev{
  S_i = \frac{1}{32\pi^2} \Tr \CW_{i,\alpha}\CW^\alpha_i\ ,
}
where $\CW_{\alpha,i}$ is the fermionic chiral superfield for
the vector multiplet of the
unbroken gauge group $U(N_i)$.
Much of the nonperturbative information about the gauge
theory is captured in the low-energy effective action for
$S_i$ \vy.

The conjectured nonperturbative
glueball superpotential is:
\eqn\dvconja{
  W(S_i) = \sum_i N_i \frac{\p F_0(S)}{\p S_i} + b^i S_i\ .
}
Here $F_0(S_i=g_s M_i)$ is the
saddle-point solution to the free energy
of the following holomorphic matrix integral:
\eqn\dvconjb{
  Z = \int d\Phi e^{-\frac{1}{g_s} W_0(\Phi)} \sim
e^{-\frac{1}{g_s^2} F_0(g_s M_i)}\ .
}
Here $\Phi$
is an $M\times M$ complex matrix.
It can be written
as $\Phi = U^\dagger \Lambda U$ where $\Lambda$ is diagonal
and $U$ is unitary.  
The $M_i$, $\sum_i M_i = M$,
are specified by running $M_i$ of the 
eigenvalue contours over the $i$th critical point of $W_0$.
In the simplest example of the most symmetric vacuum, 
the integral over $U$ factors out and
leads to the volume factor \refs{\OoguriGX,\DijkgraafPP}
\eqn\volume{
  \int dU = M^{M^2/2} = e^{\half M^2 \ln M} = 
  e^{\frac{1}{g_s^2} S^2 \ln S}
}
Therefore, the volume factor contains the
Veneziano-Yankielowicz term \refs{\vy}, from which the
existence of $N$ supersymmetric vacua \refs{\constraints}
can be understood \refs{\vy,\pesk} (for a review
of this approach see chapter 8 of \refs{\amatietal}).

\subsec{Perturbative arguments}

The perturbative argument
for this result, 
enunciated in \dvpert, 
and summarized in \refs{\cdswone,\multitrace},
is phrased entirely in terms of 
superfield perturbation theory.
%
In this language it is simple to introduce 
explicit soft SUSY breaking in the F-term sector, by treating
the couplings of the theory as (chiral) spurion superfields,
and letting the auxiliary components take nonzero
values.  
We will focus on the illustrative example of 
the superpotential
\eqn\cubicsp{
   W_0(\Phi) = \half m \Phi^2 + \frac{1}{3} g \Phi^3
}
and replacing $m$ with a chiral ``spurion'' superfield, so that:
\eqn\mtoM{
   m \longrightarrow M = m + \theta^2 \Delta
}
in \cubicsp, the classical Lagrangian is modified by a 
SUSY-breaking term which in component form is:
\eqn\softterm{
   \delta \CL = \int d^4 x \left(\Delta \tr \phi^2 + 
   \bar{\Delta} \tr \bphi^2\right)\ .
}
All of the arguments of \refs{\dvpert,\cdswone}\ 
proceed
as before; they are based on superfield
perturbation theory, and we need merely replace
$m$ with $M$ in each diagram.\foot{Ref. \grza\
does precisely this, in order
to compute the gaugino mass induced
by soft breaking.}  The result of this can be seen by
studying the leading correction
to the Veneziano-Yankielowicz superpotential:
\eqn\corrsp{
   W(S) = N S \ln (S/\Lambda_0^3) + 2\pi i \tau S + 2 N 
   \frac{g^2}{m^3} S^2 + \ldots
}
Replacing $m$ with $M$, and 
doing the superspace integral,
the correction term to the
Langrangian is linear in $\Delta$:
\eqn\corrpot{
   L = L_{(\Delta = 0)} - 6 \frac{g^2}{m^4} \bar{\Delta}
   \sigma^2 + {\rm h.c.}
}
where $S = \sigma + \sqrt{2}\theta \psi_S + \ldots$.
This is a soft mass for the glueball scalar.
This result is consistent with the 
string theory interpretation, 
where the glueball superfield 
is part of a closed-string vectormultiplet.
This multiplet is associated to a 3-cycle
which arises via a conifold transition 
from the 2-cycle which the 5-branes were 
wrapping.
We have shown earlier that 
the soft-breaking parameter $\Delta$ 
arises as a mode of the 
flux through the cycle whose period is $m$.
Using again the GVW superpotential, 
one finds a soft term 
of the form \corrsp.

Although the F-terms behave simply, the explicit breaking
ruins the holomorphicity properties of the theory.
One can see this most simply by studying a
single neutral chiral scalar multiplet $\Phi$ in four dimensions,
with canonical kinetic terms, and
with the same bare 
superpotential and soft SUSY-breaking term as above.
To zeroth order in $\Delta$, the chiral correlator
for the scalar components $\phi$ vanishes:
\eqn\chiralcorra{
   \langle \phi(x_1) \phi(x_2) \rangle = 0 + \CO(\bar{\Delta}) + \ldots
}
To first order in $\Delta, \bar{\Delta}$, the above
correlator has a nonconstant antiholomorphic piece:
\eqn\chiralcorrb{
\eqalign{
   \langle \phi(x_1) \phi(x_2) \rangle & = 
   \langle \phi(x_1) \phi(x_2) \int d^4 x \bar{\Delta} 
      \bar{\phi}^2 \rangle + \CO(|\Delta|^2,\bar{\Delta}^2)\cr
   & = 2 \bar{\Delta} G(x_1,x_2) + \ldots
}}
where $G(x,y) = \langle \phi(x)\bar{\phi}(y)\rangle$.  
We will see this more generally below,
but this simple example already reminds us that holomorphicity
of correlators of chiral operators fails when
SUSY is spontaneously or explicitly broken.

Why, then, did the superfield arguments carry through?  
The point is that we were always asking explicitly 
holomorphic questions.  Since we are treating the
SUSY-breaking parameter as a component of a nonpropagating
superfield, the superfield action is still broken up into
a D-term part integrated over $d^4 \theta$ and an F-term
part integrated over $d^2 \theta$.  
The latter will remain holomorphic in fields,
in particular in the coupling superfields.  
The former can contain terms that mimic
F-terms after expanding in $\Delta,\bar{\Delta}$ and integrating
over $d^2 \bar{\theta}$ \refs{\EvansIA}.  However
these terms are proportional to $\bar{\Delta}$ and
do not contribute in the $\bar{\Delta} \to 0$ limit.
So we may compute
the F-term by setting the antiholomorphic soft breaking parameter
$\bar{\Delta}$ to zero, and so isolating the
purely holomorphic dependence of the chiral correlators.
All of this will make sense in perturbation theory.
We will have to be more careful when making nonperturbative
arguments.

\subsec{Anomaly arguments}

In addition to the perturbative aruments in \refs{\dvpert,\cdswone},
Cachazo \etal\ have given a non-perturbative argument for the
Dijkgraaf-Vafa conjecture.  Let us outline these steps, in order to 
highlight the places that they can fail when 
supersymmetry is broken.

The first important point is that correlators of 
gauge-invariant chiral operators are independent of
position, and therefore (by cluster decomposition)
factorize.  The basis of this argument is the
simple observation \refs{\nsvza,\amatietal,\cdswone}:
\eqn\posind{
   \frac{\p}{\p x_1^\mu} \bra{0} \CO(x_1)\CO(x_2)
   \ldots \CO(x_n)\ket{0} =
   \sigma_{\mu,\alpha\dot{\alpha}} 
   \bra{0} \left\{\bar{Q}^{\dot{\alpha}},[Q^\alpha,\CO(x_1)]\right\}
   \CO(x_2)\ldots \CO(x_n)\ket{0} = 0
}
This vanishes because the operators are chiral
and so $\bar{Q}$ on both sides of the anticommutator
can be pushed through the other operators and made
to act on the vacuum.  So long as the vacuum is
supersymmetric, $\bar{Q}$ annihilates the vacuum
and the derivative vanishes.

The second important claim is that the
correlators of chiral operators are
holomorphic in the couplings. 
More precisely, if $\lambda$ is some superpotential
coupling for the term
$\int d^4 x d^2 \theta ~\lambda \CO_\lambda$, then
\eqn\antiholderiv{
   \frac{\p}{\p \bar{\lambda}}
   G(x_1,\ldots x_n) =
   \bra{0} \CO(x_1)\ldots \CO(x_n)
   \int d^4 x \left\{\bar{Q},[
   \bar{Q},\bar{\CO}_{\bar{\lambda}}]\right\}
   \ket{0}
}
is only requred to vanish when $\bar{Q}$ (anti-) commutes
with the chiral operators {\it and} annihilates the vacuum.
The proof in \refs{\cdswone}\ depends on the statement
that for some coupling $g_i$ in
$W(\Phi)$, the derivatives of the low-energy superpotential
can be written as:
\eqn\relation{
        \frac{\p W_{eff}(g_k,S,...)}{\p g_i}
        = \langle \frac{\p W(\Phi)}{\p g_i} \rangle\ .
}
This statement depends on the right hand side being a
holomorphic function of $g_i$.

The last ingredient of this proof
which requires a SUSY-invariant vacuum
involves the Konishi anomaly,
reflecting the anomalous variation of the
measure of the path integral for $\Phi$ under
the field redefinition
\eqn\redef{
   \Phi \longrightarrow \Phi + \delta\Phi = \Phi + 
   f(\Phi, \CW_\alpha)\ .
}
Explicitly, the current generating this
transformation 
\eqn\koncurr{
   J_f = \Tr \bar{\Phi} e^{V} f(\Phi,\CW_\alpha)\ ,
}
where $V$ is the real superfield for the $U(N)$ vector multiplet,
satisfies an anomalous conservation law:
\eqn\konanom{
   \bar{D}^2 J_f = \Tr f \frac{\p W(\Phi)}{\p \Phi}
   + \frac{1}{32\pi^2} \sum_{ij} \left[
   \CW_\alpha, \left[\CW^\alpha, \frac{\p f}{\p \Phi_{ij}}\right]
   \right]_{ji}
}
where the indices $i,j$ are $U(N)$ adjoint indices.
Since $\bar{D}^2 J$ can be written as a 
$\bar Q$-commutator, 
upon taking the expectation value of
\konanom, the left hand side vanishes,
if the vacuum is supersymmetric.  With a judicious
choice of $f$, and using the
factorization of chiral correlators,
Cachazo \etal\ then map the 
resulting equation
to the loop equations of the
matrix model described above.

These three points in the argument of \refs{\cdswone}\ 
fail when the vacuum is not supersymmetric.
However, if we add the soft breaking
term discussed in the previous section,
the arguments hold if we restrict ourselves to the
parts of the chiral correlators that depend solely on the
holomorphic coupling $\Delta$ and not on the
antiholomorphic coupling $\bar{\Delta}$.  The point is that
if we promote an F-term coupling to a superfield
whose auxiliary component has vev $\Delta$, the perturbation
to the Lagrangian taks the form:
\eqn\generalpert{
   \delta L = \int d^4 x \left(\Delta p(\phi) + \bar{\Delta}
   \bar{p}(\bar{\phi})\right)
}
where $\phi$ denotes the bottom components of any of the chiral
superfields that appeared in this perturbation.
Chiral correlators 
\eqn\undefcorr{
   G(x_1,\ldots,x_n) = \langle \CO(x_1)\ldots\CO(x_n)\rangle
}
are deformed to:
\eqn\defcorr{
   G_{\Delta,\bar{\Delta}} = 
   \langle \CO(x_1)\ldots\CO(x_n) e^{ - 
   \left( \int \Delta p(\phi) - \int \bD \bar{p}(\bar{\phi})\right)
   }\rangle~.
}
In general there will be a non-trivial dependence
on both $\Delta$ and $\bD$.  However, if we expand
$\exp\left\{-\int\left(\Delta p(\phi)+
\bD\bar{p}(\bar{\phi})\right)
\right\}$ in $\bD$ and keep the $\bD$-independent term,
the resulting correlator can be written
as a correlator of chiral operators in the original
vacuum,\foot{
The vacuum will of course be modified by the 
addition of the soft-breaking terms. 
However, the following argument
shows that 
this does not affect the 
$\bar \Delta$-independent 
part of chiral correlators.
The perturbed vacuum can
be calculated perturbatively in $\Delta,\bD$ about
the SUSY vacuum.  Since, as we will argue below, 
this perturbation series is expected to be
analytic when the SUSY breaking is soft, 
we can isolate the terms which are independent of $\bD$.
These terms will be states which are connected to the
unperturbed 
vacuum by chiral operators, and the anomaly argument goes through.
}
for which the above anomaly-based arguments apply.

This argument works to all
orders in a perturbation expansion in the soft coupling.
One might worry that nonperturbative terms in $\bD$
would make it difficult to separate out the holomorphic part
of a correlator.  However, in general one does not expect 
nonperturbative behavior in the coefficients of
relevant operators (which is precisely what 
characterizes these soft terms), 
as long as they do not change
the large-field behavior of the action.  Non-analytic
behavior in a coupling occurs when the vev of the
operator it multiplies 
is comparable to the vevs of other terms in the action.
If there is a $\phi^4$ term
in the potential energy in addition to the soft breaking term,
then this marginal 
term dominates over the soft-breaking 
term in the action.
Said another way,
one may try to apply Dyson's argument \refs{\dysonPT} 
for the breakdown of perturbation
theory at large orders.  Applied to the coupling
which dominates at large field values -- \eg\ the quartic
scalar coupling in a renormalizable field theory --
it states that since the theory is unstable for negative
values of the coupling, it must have a singularity when the 
coupling vanishes.  Therefore the perturbation series diverges.
But if one flips the sign of a $m^2$ term in the presence of a quartic
coupling, physics is not singular and so perturbation
theory in a soft mass $(\delta m)^2$ should converge
(there is potentially an infrared divergence,
but we are assuming the unperturbed 
supersymmetric mass of $\Phi$ is 
nonzero).\foot{We thank S. Shenker for
a discussion on this point, and for reminding us of Dyson's argument.}

One {\it will} have to worry if the
explicit SUSY breaking terms 
include marginal or large irrelevant couplings that change
the asymptotics of field space.  At this point
the correlators can be nonperturbative in the
SUSY-breaking parameters and one cannot meaningfully
extract the holomorphic piece.  This should not
come as a surprise, as the ultraviolet theory
will depend strongly on the SUSY-breaking parameters
in such a case.

\newsec{Conclusion}


In this paper, we have made a 
stringy identification 
of the auxiliary fields of $\CN=2$ multiplets in type II on a CY.
This identification 
extends our knowledge 
of bosonic couplings between sectors of string theory 
to $\CN=2$ multiplets.
Specifically, 
closed-string multiplets act as spurion superfields 
for open-string modes,
and, in a local model,
spurions for localized closed-string modes
are closed-string fields which 
are not normalizable.  
This latter fact fits nicely 
with the modularity of physics
made possible by these  
constructions.
The couplings of the localized modes
performing some service to physics
are determined by closed-string modes 
which also have support far 
away in the CY;
it is the vevs of auxiliary fields for these modes
that transmit SUSY-breaking between modules.

Our discussion has covered 
what should be a 'fundamental domain'
for the action of mirror symmetry (to the extent that
mirror symmetry is a relevant concept in the presence
of magnetic fluxes).
The set of objects which can arise as 
auxiliary fields in this context 
is the same as the set of objects which 
can appear as central charges 
of the supersymmetry algebra.
Heuristically, this is because 
they both arise by 
acting with two supercharges on
the lowest component of a supermultiplet.

A fact which is highlighted by our work 
is that in type II string theory, we do not understand 
precisely what all of these objects are. In particular,
we have been able to understand auxiliary fields
in the vector multiplets in
type IIB string theory, and auxiliary fields
in the hypermultiplets in type IIA string theory.
Both involve NS-NS three-form flux.  
In addition to 
RR and NSNS fluxes, 
NUT charges \nutcharge\
can also 
play the role of auxiliary fields for 
vectors in type IIA and hypers in type IIB 
\KachruSK.
The value of such a charge can 
be thought of, at least heuristically, 
as a number of KK monopoles, which are 
related by U-duality to 
D-branes and NS-branes.

Still, the {\it generic} object which 
can play the role of a central charge
in type II string theory is not understood.
This becomes clear in a semi-flat approximation
to the geometry as a flat $T^3$ fibration \syz.
This approxiation is good near a large complex structure point.
In this description, the central charges 
are encoded in the monodromies acting on the string theory 
on the $T^3$ fibers, associated to 
homotopy generators of the base.
For example, the fact that $N$ units of flux $F$ 
are supported near a singular fiber 
is encoded by the fact that 
the corresponding potential $C$ ($F = dC$)
undergoes a shift through $N$ periods $ C \to C + 2\pi N$
during a tour around that point in the base.

In this approximation, mirror symmetry is 
visible as an element of the 
duality group of the $T^3$ fibers.
Conjugating the monodromy group 
of a generic IIB solution 
with only RR and NSNS fluxes by this element 
leads to a set of central charges which 
are not merely curvatures and fluxes.
The monodromy group will generically 
include non-geometric 
elements of the U-duality group 
(such as T-dualities),
and even non-perturbative ones
(such as S-dualities).
Such string backgrounds, studied in \eg\ 
\refs{\nongeometric},
can often be considered resolutions of asymmetric orbifolds.
In this sense, the mirror of even 
the deformed conifold, with NSNS 3-form flux through 
the 3-sphere, is non-geometric \simeon.

The identification of 
the relevant central charges would be
useful at least because it would help in 
identifying the domain walls 
which change their values.  For the case of RR and NSNS flux vacua,
D5- and NS5-branes on holomorphic 
curves provide BPS domain walls between vacua
with different values of RR and NSNS fluxes
\refs{\GVW}, and non-BPS domain walls between
vacua with different amounts of supersymmetry
\refs{\KachruNS}. 
As an example of the utility of these domain walls, 
the fact that they interpolate between SUSY vacua
with different fluxes
provides an alternative derivation 
of the GVW superpotential \refs{\GVW}.
It would be quite useful to understand microscopically 
the domain walls between vacua with different values 
of the other kinds of central charges.

We hope to shed
some light on these questions in upcoming work \refs{\shaljm}.


\vskip .6cm
\centerline{\bf Acknowledgements}

We would like to thank A. Adams, N. Arkani-Hamed, 
M. Berkooz, N. Berkovits, 
O. DeWolfe, S. Hellerman, 
S. Kachru, S. Katz, M. Peskin, 
L. Randall, H. Schnitzer, 
M. Schulz, S. Shenker, R. Tatar, T. Taylor and N. Wyllard for
helpful discussions and comments.  The research of
A. Lawrence is supported in part by NSF grant PHY-0331516.  
J. McGreevy
is supported by a Princeton University Dicke Fellowship, 
and by the Department of Energy under 
Grant No. DE-FG03-92ER40701.

A.L. would like to thank the SLAC Theory Group
and the Stanford ITP for their hospitality during an
early stage of this project.  He would also like to thank
the Institute for Studies in Theoretical Physics
and Mathematics (IPM) in Tehran, and especially Shirin
Davarpanah and Maryam Soltani, for running
the ISS2003 school in Anzali, Iran so smoothly that it
was actually possible for the organizers to get work done.
Finally, he would like to thank the Aspen Center for Physics and
the Kavli Institute for Theoretical Physics at UCSB,
where much of this work was carried out. 

\listrefs
\end

We would like to find a deformation of the
string theory background that corresponds to
the auxiliary component $F_a$.
This can be derived from the equation of motion for
the auxiliary field:
\eqn\auxeqn{
   \bar{F}_a = \frac{\p W(t^a)}{\p t^a}\ .
}
We take $W$ to be that given in \GVWsuperpot.
Combined with \perderiv, we find that:
\eqn\auxanswer{
   \bar{F}_a = \int_X G \wedge \omega_a
}
In other words, $\bar{F}$ corresponds to 
the $(1,2)$ component of $G$, which we know
breaks SUSY.

If this is the correct auxiliary field, there should
be a quadratic mass-like term
for these components of the flux:
\eqn\ftermfull{
   L = \sum_{a} g^{a,\bar{a}}\left(F_a F_{\bar{a}} -  
   F_{{\bar a}} \p_a W - F_a \p_{\bar{a}} \bar{W}\right) + \ldots
}
Indeed, such a term arises from the 10d kinetic energy for
the NS-NS and RR two-form potentials:
\eqn\kinetic{
   S_{G} = \int_{M^{10}} F \wedge {}^\ast F + \ldots
}
Upon reducing to four dimensions, there will be a term
with indices for $F$ and $G$ living entirely in $X$.

The SUSY-breaking flux is a $(1,2)$ form
through a linear combination of this
three-cycle and the dual three-cycle.  To see this,
again use the equation:
\eqn\auxexample{
\eqalign{
   \bar{F}_a & = \p_a W(t) = \p_a \sum_b \left[ \int_{A^b} G
      \int_{B_b} \Omega(t) - \int_{B_b} G \int_{A^b} \Omega \right]\cr
      & = \p_a \sum_b \left[ \CF_b \int_{A^b} G - t^b \int_{B_b} G
         \right] \cr
      & = \sum_b \CF_{b,a} \int_{A^b} G - \int_{B_a} G\ 
}}
where $\CF$ is the prepotential and $\CF_a \equiv \p_a \CF$ is
the dual period.  
In the present case we take $A^a$ to be the noncompact cycle $A$
and $B_a$ to be its dual.
The last line vanishes identically if $G$
is a $(2,1)$ form, proving the statement above.
</PRE></BODY></HTML>
